\documentclass[twocolumn,
aps,prd,   
               preprintnumbers,numbers,sort&compress,
               nofootinbib,
                            showpacs,
               colorlinks,
               linkcolor=blue,
               citecolor=blue]{revtex4-1} 

\usepackage{graphicx}
\usepackage{dcolumn}
\usepackage{bm}
\usepackage{hyperref}

\usepackage{graphicx}
\usepackage{dcolumn}
\usepackage{bm}
\usepackage{hyperref}

\usepackage{amsmath}
\usepackage[caption=false]{subfig}
\usepackage{gensymb}

\def\ra{\rangle}
\def\la{\langle}
\def\rmd{\mathrm{d}}  

\newcommand{\be}{\begin{eqnarray}}
\newcommand{\ee}{\end{eqnarray}}
\newcommand{\beq}{\begin{equation}}
\newcommand{\eeq}{\end{equation}}
\newcommand{\kmps}{\,\mathrm{km}\,\mathrm{s}^{-1}} 
\newcommand{\exclude}[1]{}

\newcommand{\jcap}{JCAP}
 \exclude{
 \newcommand{\prl}{Phys. Rev. Lett.}

\newcommand{\nat}{Nature}
\newcommand{\prd}{Phys. Rev. D}

}
\graphicspath{{./Figures/}} 

\begin{document}
       \title{Multi-Modal Clustering  Events observed by Horizon-10T  and    Axion Quark Nuggets }
       \author{Ariel  Zhitnitsky}
       \affiliation{Department of Physics and Astronomy, University of British Columbia, Vancouver, V6T 1Z1, BC, Canada}
     
      \begin{abstract}
        The Horizon-10T collaboration  \cite{Beisembaev:2016cyg,2017EPJWC.14514001B,Beznosko:2019cI,2019EPJWC.20806002B,Beisembaev:2019nzd} have reported observation of Multi-Modal    Events (MME)     containing multiple peaks suggesting their  clustering origin. These events are proven  to be hard
        to explain in terms of conventional cosmic rays (CR). We propose that these MMEs   might be result of the 
    dark matter annihilation events within  the so-called axion quark nugget (AQN) dark matter 
    model, which was originally invented for completely different purpose to explain the observed  similarity between the dark   and the visible components  in the Universe, i.e. $\Omega_{\rm DM}\sim \Omega_{\rm visible}$ without any fitting parameters.  We  support this proposal by demonstrating that  the  observations  \cite{Beisembaev:2016cyg,2017EPJWC.14514001B,Beznosko:2019cI,2019EPJWC.20806002B,Beisembaev:2019nzd}, including the    frequency of appearance, intensity, the spatial distribution, the  time duration, the clustering features,  and many other properties nicely match the emission characteristics   of the   AQN annihilation events in atmosphere.  We list a number of features  of the AQN events   which   are very distinct  from conventional CR air showers. The observation (non-observation) of these features  may  substantiate  (refute)  our proposal. 
        \end{abstract}
     
       \maketitle

     \maketitle
\section{Introduction}\label{sec:introduction}
 In this work we discuss two naively unrelated stories.    
The  first one is  the study  of  a specific dark matter (DM) model, the  so-called axion quark nugget (AQN) model  \cite{Zhitnitsky:2002qa}, see a brief overview of this model below. The second one deals
with  the recent puzzling  observations  \cite{Beisembaev:2016cyg,2017EPJWC.14514001B,2019EPJWC.20806002B,Beznosko:2019cI,Beisembaev:2019nzd} by the Horizon-10T (H10T) collaboration  of the Multi-Modal    Events (MME). 
     We overview the corresponding events  below in details.  
  We also highlight some difficulties in interpretation of these events in terms of the standard  CR air showers. The unusual  features of MME include:
  
  {\it {\bf 1}. ``clustering puzzle":} Two or more peaks separated by $\sim 10^2$   ns are present in several  detection  points, while entire event may last $\sim 10^3$ ns. It can be viewed as many fronts separated by $\sim (10^2-10^3)$   ns, instead of a single  front; 
  
   {\it {\bf 2}. ``particle density puzzle":} The number density of particles recorded at different  detection  points apparently weakly  dependent  on 
   distance from  Extensive Air Showers (EAS) axis;
   
      {\it {\bf 3}. ``pulse width puzzle":} The width of each individual pulse is around $(20-35)$ ns and apparently does not   depend  on  distance from    EAS axis;

   {\it {\bf 4}. ``intensity puzzle":} The observed intensity of the events (measured in units of a number of particles per unit area) is of order $\rho\sim (100-300) \rm m^{-2}$
   when measured at distances $(300-800)$ m from EAS axis. Such intensity would   correspond to   the CR energy of the primary particle on the level $E_{p}\gtrsim 10^{19}$ eV which would have  dramatically different event rate in comparison with observed rate recorded by H10T. 
  
  Before we  proceed with our explanation of the proposal to view    the MMEs   as the AQN events one should mention that 
  similar  unusual features of the CR air showers have been noticed long ago for the first time by  Jelley and Whitehouse \cite{Jelley_1953} in 1953. Later, EAS exhibiting the unusual time structures were studied by several independent experiments, see e.g. \cite{PhysRev.128.2384}. We refer to ref. \cite{2017EPJWC.14514001B}  for overview and references  of the older observations and studies of such unusual events.
  
  The  considerable recent advances  in rising the resolution (on the level of a few  ns) has allowed the  H10T collaboration   dramatically improve the collection and analysis of the MMEs. 
  In particular, during $\sim 3500$ hours of operation the  H10T collaboration  collected more than $10^3$ MMEs.

  In what follows  we overview   the  observation  \cite{Beisembaev:2016cyg,2017EPJWC.14514001B,Beznosko:2019cI,2019EPJWC.20806002B,Beisembaev:2019nzd}  by emphasizing  the  puzzling nature of these events  if   interpreted  as  conventional EAS events.  At the same time    the same observations  can be explained in very natural way     if    interpreted  in terms of  the AQN annihilation events as we shall argue in this work. One should mention here that a similar   conclusion   has been also reached  for  a different type of unusual CR-like events.  First,    it has been argued  in    \cite{Zhitnitsky:2020shd,Liang:2021wjx} that   the Telescope Array (TA)   
  ``mysterious bursts"  \cite{Abbasi:2017rvx,Okuda_2019} can be naturally  interpreted as  the AQN events.  Secondly, it has been also argued in \cite{Liang:2021rnv} that 
      the  Antarctic Impulse Transient Antenna (\textsc{ANITA})    observation  \cite{Gorham:2016zah,Gorham:2018ydl,ANITA:2020gmv}   of two anomalous   events    with noninverted polarity might be also a consequence of the AQN annihilation events. Important  comment here that in all those cases    the basic parameters of the AQN model remain the same as they have been fixed long ago from dramatically different observations in a very different context.  
      
      Our presentation is organized as follows.  In next Section \ref{confronts} we explain why  the  observation  \cite{Beisembaev:2016cyg,2017EPJWC.14514001B,Beznosko:2019cI,2019EPJWC.20806002B,Beisembaev:2019nzd}  listed as  {\bf 1-4} {\it puzzles} are very mysterious   events  if   interpreted  as  conventional EAS events.
In  Section \ref{AQN}  we give a  brief overview of the AQN framework with emphasize on the key elements relevant for the present studies.
In Section \ref{proposal} we formulate    our proposal on identification of the unusual Multi-Modal Events with the  upward moving AQN events.
In Section \ref{sec:rate} we estimate the event rate.  Our main  Section  is \ref{timing} where we   estimate a variety of relevant time scales and the intensity of the events. In the same section we also    confront our proposal with observations and  argue   that puzzles  {\bf 1-4} can be naturally understood  within the AQN framework. Section \ref{conclusion} is our conclusion where we suggest several tests which may   substantiate or refute our proposal  on identification of the  Multi-Modal Cluster Events with the  upward moving AQNs.

 \section{Conventional CR picture confronts  the   MME observations \cite{Beisembaev:2016cyg,2017EPJWC.14514001B,Beznosko:2019cI,2019EPJWC.20806002B}}\label{confronts}
    \begin{figure}[h]
	\centering
	\captionsetup{justification=raggedright}
	\includegraphics[width=1.1\linewidth]{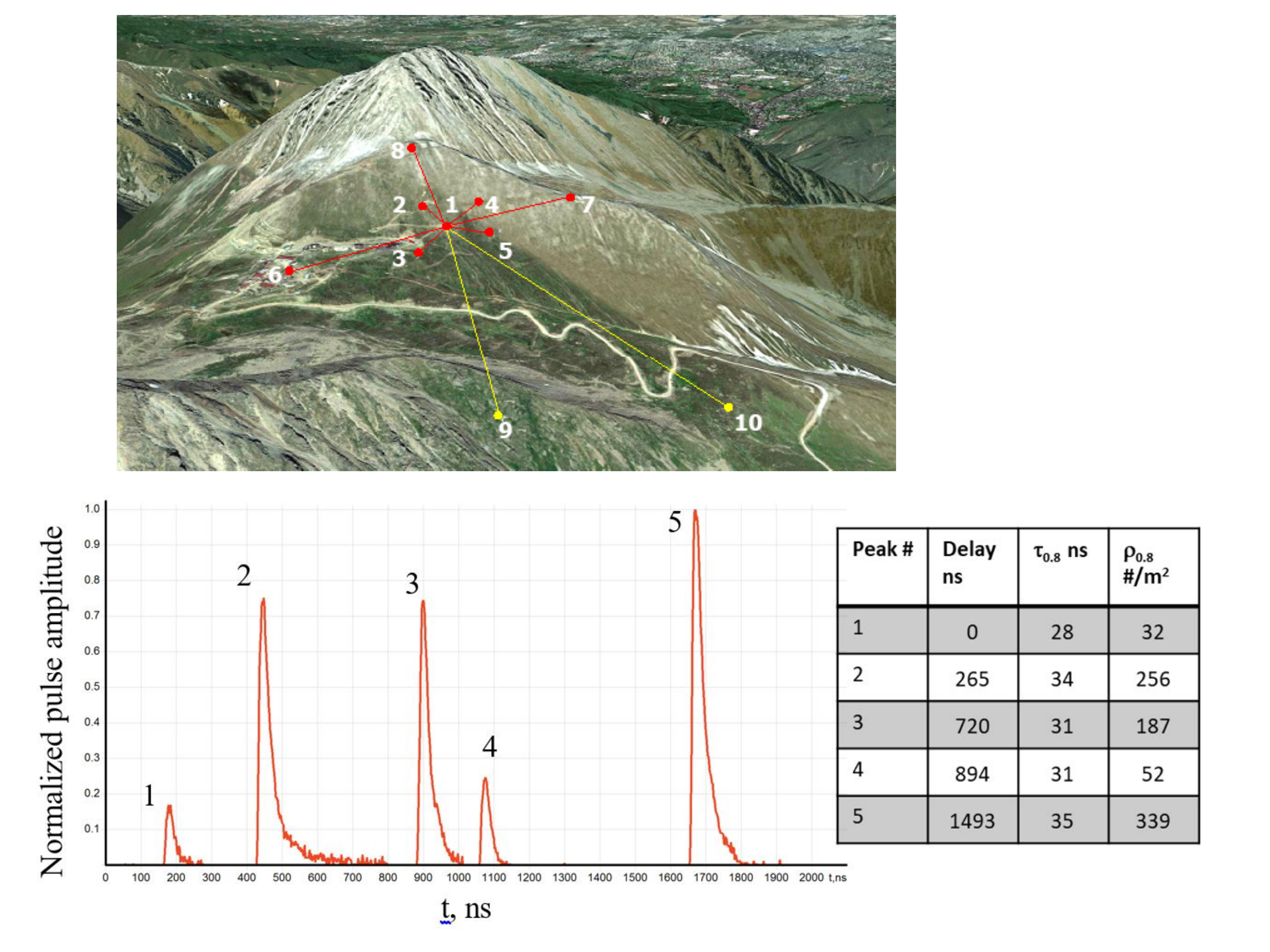}
	\caption{top: Aerial view and geometry of H10T instrument   with location of 10 detectors, adopted from \cite{Beznosko:2019cI};
	bottom: A typical MME event recorded  on April 6.2018 by H10T instrument at point $\#9$, adopted from \cite{Beznosko:2019cI}. All pulses are recorded at a single detection point. Delay times,  the width of each peak $\tau_{0.8}$ in ns, and the particle density $\rho_{0.8}$  per  $\rm m^{-2}$ within $\tau_{0.8}$ are also shown in the table. }.  
	\label{pulses}
\end{figure}
 \begin{figure}[h]
	\centering
	\captionsetup{justification=raggedright}
	\includegraphics[width=1.1\linewidth]{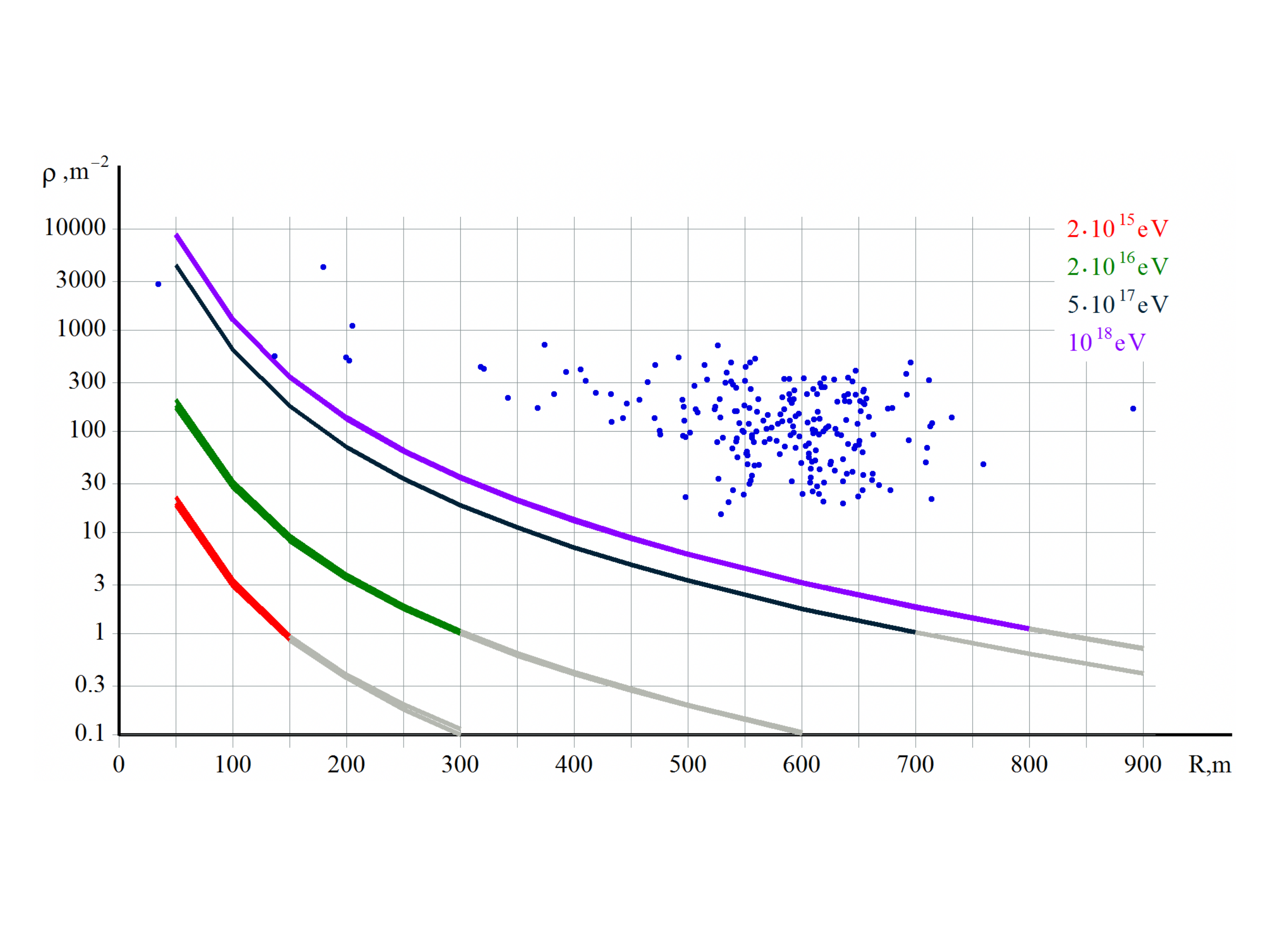}
	\caption{ Solid lines: the particle density distribution $\rho(R)$ in simulated EAS disk versus  distance from axis for different energies shown by different colours, depending on energy of the CR. Blue dots: cumulative particle density for each bimodal pulse vs. distance to EAS axis, adopted from \cite{Beznosko:2019cI}.}.  
	\label{density}
\end{figure}
 \begin{figure}[h]
	\centering
	\captionsetup{justification=raggedright}
	\includegraphics[width=1.1\linewidth]{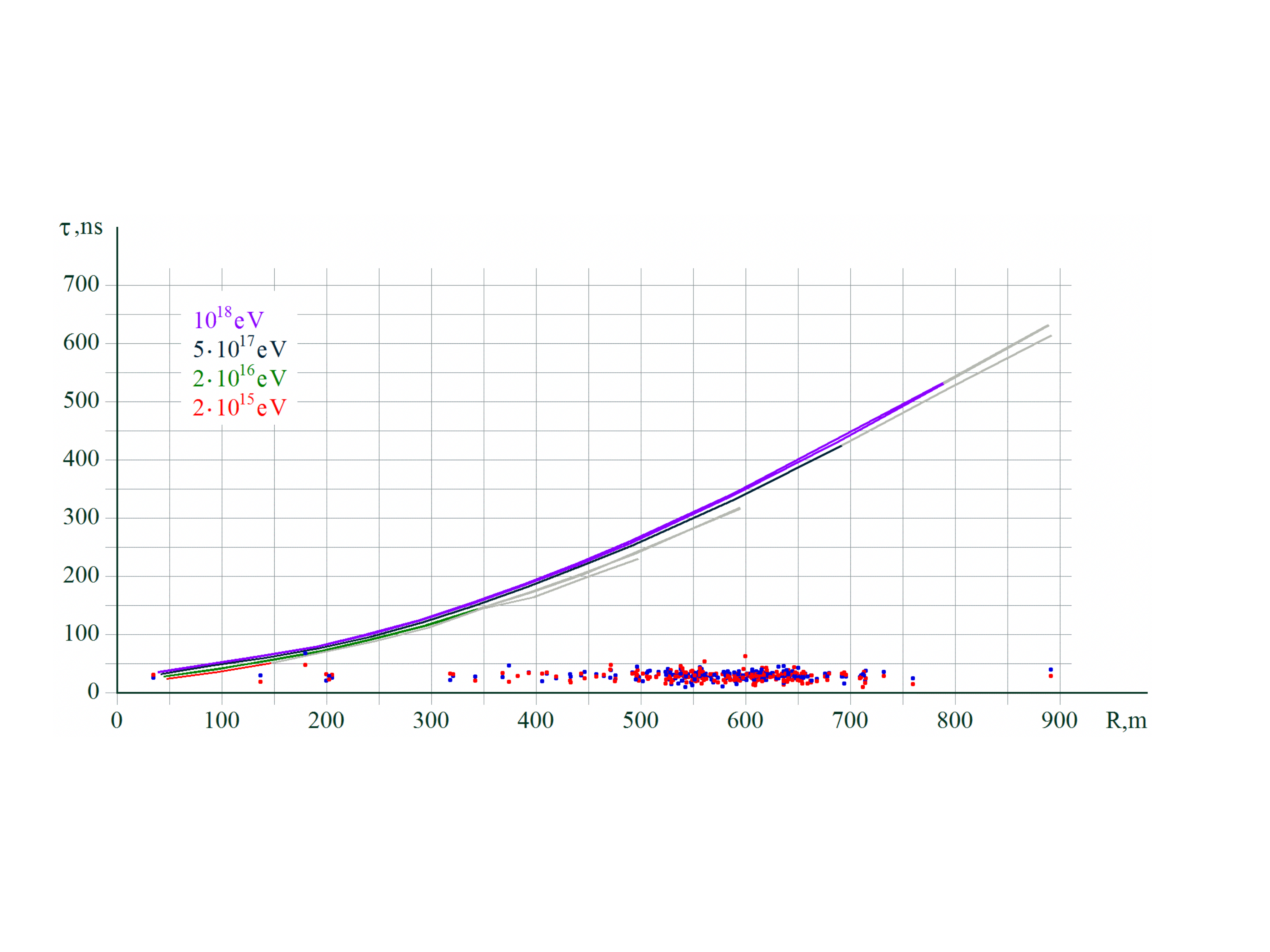}
	\caption{ Solid lines:  simulated EAS disk width versus  distance from axis for different  energies shown by different colours.    Pulse width $\tau_{0.8}$ in ns  for the first pulse (in blue) and second pulse (in red) for the bimodal pulses versus  distance to the EAS axis, adopted from \cite{Beznosko:2019cI}.}.  
	\label{width}
\end{figure}

 This section is devoted to the first part of our story where we describe the MME observations and argue why the observed  events are inconsistent with conventional CR interpretation, while the next section \ref{AQN} is devoted to the second part of our story, the AQN dark matter model. 
 
 We start by reviewing the conventional picture of the EAS. It is normally assumed that EAS can be thought as a disk -pancake with well defined EAS axis.
 It is also assumed that EAS represents an uniform, without any breaks structure. The standard picture also assumes that   the particle density drops smoothly with distance when moving away from the core, while the thickness of the EAS pancake  increases with the distance from the core. Now we want to see why this conventional picture is in dramatic contradiction with   observations of the MME events.
 
{\bf 1.}  Indeed, a typical MME is shown on Fig. \ref{pulses} where a complicated 
temporal features are explicitly seen. Several peaks separated by $\sim 10^2$   ns in a single detector represent  the {\it ``clustering puzzle"}, listed  above.
In conventional EAS   picture one should see a single pulse in each given detector with the amplitude which   depends on the distance from the EAS axis. It is not what actually observed by H10T.

{\bf 2.} The manifestation of  the {\it ``particle density puzzle"} is as follows. On Fig. \ref{density} we show the particle density distribution $\rho(R)$ in simulated EAS disk versus  distance from axis for different energies shown by solid lines with  different colours, depending on energy of the CR. In particular, for energy of the primary particle on the level of $10^{17}$ eV one should expect  a strong  suppression $\sim 10^3$  when distance $R$ to the EAS axis changes from $R\simeq 100$ m to $R\simeq 700$ m. It is not what actually observed by H10T: the density of the particles $\rho(R)$ is not very sensitive to the distance to the EAS axis, and remains essentially flat for the entire region of observations. Furthermore, the magnitude of the density $\rho(R)$ is much higher than it is normally expected for CR energies $\sim (10^{17}-10^{18})$ eV.

{\bf 3}. The manifestation  of the {\it ``pulse width puzzle"} is  as follows. On Fig. \ref{width}  we show the   simulated EAS disk width versus  distance from axis for different  energies shown by solid lines in different colours. As explained above  the thickness of the EAS pancake  increases with the distance from the core. Therefore, the pulse width must  also increase correspondingly as shown by sold lines on Fig. \ref{width} for different energies. It is not what actually observed by H10T: all MME events show 
similar duration of the pulse width on the level of $(20-35)$ ns irrespective  to  the distance to the AES axis. This observation is in dramatic conflict with conventional picture as outlined above. 

{\bf 4}.   The manifestation  of the  {\it ``intensity puzzle"} is  as follows.  The   Fig. \ref{density} suggests that the charged particle  density  $\rho(R)$ varies between $(30-300) $ particles per $m^2$ at the distances $(500-700)$ m from the EAS axis. This is at least factor $10^2$ above the expected  $\rho(R)$ for the primary particle  with energy in the interval  $(10^{17}-10^{18})$ eV. Only the particles  with energies well above $10^{19}$ eV could generate such enormous particle  density as shown on Fig. \ref{density}. However the frequency of appearance  for such highly energetic particles is  only once every few years. This represents the {\it ``intensity puzzle"} when the intensity of the events 
  estimated from $\rho(R)$  for MME  is several  orders of magnitude higher than the energy estimated by the event rate. This shows  a dramatic inconsistency between the {\it measured} intensity and {\it observed} event rate carried out by  one and the same detector. 
 
  We conclude this section with the following comment. From the  puzzles listed above it must be obvious that the events detected by H10T are not the conventional CR air showers as they demonstrate enormous inconsistency with standard CR interpretation. What it could be? Before we put forward our proposal   we would like to briefly overview in next section \ref{AQN}  the second part of our story- the AQN   dark matter model.

 \section{The AQN   Dark matter model }\label{AQN}
 We start with few historical remarks and motivation of the AQN model in subsection \ref{basics},  while  in subsection 
 \ref{earth} we overview recent observations of CR-like events (such as puzzling  bursts observed by the TA experiment and ANITA's   two anomalous events with non-inverted polarity) of some mysterious events which could be explained by the AQN events hitting the Earth. Finally, in section \ref{internal} we briefly overview some  specific features of the AQNs  traversing the Earth (such as internal temperature, level  of ionization, etc).  These characteristics  will be important  for the present study  interpreting the MMEs   as the  AQN events.  
 
 \subsection{The basics}\label{basics}
   The   AQN dark matter  model  \cite{Zhitnitsky:2002qa} was  invented long ago with a single motivation 
 to   explain in natural way the observed  similarity between the dark matter and the visible densities in the Universe, i.e. $\Omega_{\rm DM}\sim \Omega_{\rm visible}$ without any fitting parameters. We refer to recent brief review  \cite{Zhitnitsky:2021iwg} on the AQN model. 
 Here we want to mention few key elements which are important for this work.
 
 The AQN construction in many respects is 
similar to the Witten's quark nuggets, see  \cite{Witten:1984rs,Farhi:1984qu,DeRujula:1984axn},  and    review \cite{Madsen:1998uh}. This type of DM  is ``cosmologically dark'' not because of the weakness of the AQN interactions, but due to their small cross-section-to-mass ratio, which scales down many observable consequences of an otherwise strongly-interacting DM candidate. 

There are two additional elements in the  AQN model compared to the original models \cite{Witten:1984rs,Farhi:1984qu,DeRujula:1984axn,Madsen:1998uh}. First new element is the presence of the  axion domain walls which   are copiously produced  during the  QCD  transition\footnote{The axion field had been introduced into the theory to resolve the so-called the strong ${\cal CP}$ problem which  is related to the fundamental initial parameter $\theta_0\neq 0$.
  This source of ${\cal CP}$ violation  is no longer 
available at the present time  as a result of the axion's  dynamics in early Universe. 
One should mention that the axion  remains the most compelling resolution of the strong ${\cal CP}$ problem, see original papers 
 on the axion \cite{axion1,axion2,axion3,KSVZ1,KSVZ2,DFSZ1,DFSZ2}, and   recent reviews \cite{vanBibber:2006rb, Asztalos:2006kz,Sikivie:2008,Raffelt:2006cw,Sikivie:2009fv,Rosenberg:2015kxa,Marsh:2015xka,Graham:2015ouw, Irastorza:2018dyq}.}. The axion field $\theta (x)$ plays a dual role in this framework: first  it serves  as  an   additional stabilization factor for the nuggets,     which helps to alleviate a number of  problems with the original nugget construction  \cite{Witten:1984rs,Farhi:1984qu,DeRujula:1984axn,Madsen:1998uh}.  Secondly, the same axion field $\theta (x)$ generates the strong and coherent $\cal{CP}$ violation in the entire visible Universe. 

 This is because the $\theta (x)$ axion field before the QCD epoch could be thought as   classical $\cal{CP}$ violating field correlated on the scale of the entire Universe.
The axion field starts to oscillate at the QCD transition by emitting the propagating axions. However, these oscillations remain coherent on the scale     of the entire Universe. Therefore, the $\cal{CP}$ violating phase remains coherent on the same enormous scale.

  Another feature of the  AQN model which plays
absolutely crucial role for  the present work  is that nuggets can be made of {\it matter} as well as {\it antimatter} during the QCD transition. 
Precisely the coherence of the $\cal{CP}$ violating field on large scale mentioned above provides a preferential production of one species of nuggets made of {\it antimatter}   over another 
species made of {\it matter}. The preference    is determined by the initial sign of the $\theta$ field when the formation of the AQN nuggets starts.  
The direct consequence of this feature along with coherent $\cal{CP}$ violation in entire Universe  is that  the DM density, $\Omega_{\rm DM}$, and the visible    density, $\Omega_{\rm visible}$, will automatically assume the  same order of magnitude densities  $\Omega_{\rm DM}\sim \Omega_{\rm visible}$ without any fine tuning as they both proportional to one and the same dimensional parameter $\Lambda_{\rm QCD}$. We refer to the original papers   \cite{Liang:2016tqc,Ge:2017ttc,Ge:2017idw,Ge:2019voa} devoted to the specific questions  related to the nugget's formation, generation of the baryon asymmetry, and 
survival   pattern of the nuggets during the evolution in  early Universe with its unfriendly environment, see also brief review  \cite{Zhitnitsky:2021iwg} covering these topics.

One should emphasize that AQNs are absolutely stable configurations on cosmological scales. Furthermore, the antimatter which is hidden  in form of the very dense nuggets is unavailable for annihilation unless the AQNs hit the stars or the planets. 

However, when the AQNs hit  the stars or the planets it may lead to observable phenomena. 
\exclude{In particular,  the deposition of the energy due to the AQNs hitting  the Sun 
  may explain\footnote{In fact, to resolve this problem  Parker conjectured  long ago  \cite{Parker} that ``nanoflares"   are   identified with the  annihilation events in the AQN framework.
    The luminosity  of the Extreme UV (EUV) radiation from corona due to these annihilation events is unambiguously determined by the DM density. It is very nontrivial consistency check that  the computed  luminosity from the corona nicely matches with observed EUV radiation.     The same   events  of annihilation are  also   manifested themselves as 
     the radio impulsive events in quiet solar corona as  recently recorded    by Murchison Widefield Array Observatory \cite{Mondal-2020}, which also represents a very nontrivial consistency check of the proposal  \cite{Zhitnitsky:2017rop,Raza:2018gpb} on the ``Solar Corona Mystery" resolution and accompanying radio impulsive events in quiet solar corona \cite{Ge:2020xvf}.}  the ``Solar Corona heating problem" as advocated in \cite{Zhitnitsky:2017rop,Raza:2018gpb,Ge:2020xvf}.   
 There are also very rare events of annihilation in the center of the galaxy, which, in fact, may explain some observed galactic excess emissions in different frequency bands, including famous 511 keV line. 
 }
 The strongest direct detection limit\footnote{Non-detection of etching tracks in ancient mica gives another indirect constraint on the flux of   dark matter nuggets with mass $M< 55$g   \cite{Jacobs:2014yca}. This constraint is based on assumption that all nuggets have the same mass, which is not the case  as we discuss below.
The nuggets with small masses represent a tiny portion of all nuggets in this model, such that this constraint is easily  satisfied with any reasonable nugget's size distribution.} is  set by the IceCube Observatory's,  see Appendix A in \cite{Lawson:2019cvy}:
\be
\label{direct}
\la B \ra > 3\cdot 10^{24} ~~~[{\rm direct ~ (non)detection ~constraint]},
\ee
where we formulated the constraints in term of the AQN's baryon charge $B$ rather than in terms of its mass $M\approx m_pB$.
Similar limits are   also obtainable 
from the   ANITA 
  and from  geothermal constraints which are also consistent with (\ref{direct}) as estimated in \cite{Gorham:2012hy}.
  
  \exclude{It has been also argued in \cite{Gorham:2015rfa} that that AQNs producing a significant neutrino flux 
in the 20-50 MeV range cannot account for more than 20$\%$ of the dark matter 
density. However, the estimates \cite{Gorham:2015rfa} were based on assumption that the neutrino spectrum is similar to  the one which is observed in 
conventional baryon-antibaryon annihilation events which typically produce a large number of pions and  muons and thus generate a 
significant number of neutrinos and antineutrinos in the 20-50 MeV range where 
SuperK has a high sensitivity. However, the critical difference in the case of AQNs  is that the annihilation proceeds within the 
colour superconducting (CS) phase where the energetics are drastically  different \cite{Lawson:2015cla}. The main point is that, in most CS phases, the lightest pseudo Goldstone mesons 
(the pions and kaons) have masses in the  20 MeV range, 
rather than 140 MeV in hadronic phase. This dramatically changes 
entire spectrum such that the  main assumption of \cite{Gorham:2015rfa} on similarity of the neutrino's spectrum in both phases is incorrect. The  resulting flux computed in \cite{Lawson:2015cla} is perfectly consistent with observations. Furthermore, precisely these low energy ($\lesssim 20~ \rm MeV$) AQN-induced neutrinos produced in the Earth's interior might be responsible  for explanation of the long standing puzzle of the DAMA/LIBRA  observation  of the annual modulation at $9.5\sigma$ confidence level as argued in \cite{Zhitnitsky:2019tbh}. 

The authors of ref.\cite{SinghSidhu:2020cxw} considered a generic constraints for the nuggets made of antimatter (ignoring all essential  specifics of the AQN model such as quark matter  CS phase of the nugget's core). Our constraints (\ref{direct}) are consistent with their findings including the Cosmic Microwave Background (CMB), Big Bang Nucleosynthesis (BBN), and others, except the constraints derived from    the so-called ``Human Detectors". 
As explained in \cite{Ge:2020xvf}
  the corresponding estimates of ref.\cite{SinghSidhu:2020cxw} are   oversimplified   and do not have the same status as those derived from CMB or BBN constraints\footnote{In particular, the rate of energy deposition was estimated in \cite{SinghSidhu:2020cxw} assuming that the annihilation processes between antimatter nuggets and baryons are similar to $p\bar{p}$ annihilation process. It is known that it cannot be the case because the annihilating objects have drastically different internal structures (hadronic phase versus CS phase). It has been also assumed in \cite{SinghSidhu:2020cxw} that  a typical x-ray energy  is around  1 keV, which is  much lower than direct computations in the AQN model would  suggest \cite{Budker:2020mqk}. Higher energy x-rays have much longer mean-free path, which implies that the dominant portion of the energy will be deposited outside the human body. Finally, the authors of ref.  \cite{SinghSidhu:2020cxw} assume that an antimatter nugget will result in ``injury   similar to a gunshot". It is obviously a wrong picture as the size of a typical nugget is only $R\sim 10^{-5} {\rm cm}$ while the most of the energy is deposited in form of the x-rays on centimeter  scales \cite{Budker:2020mqk} without making a large hole similar to  bullet as assumed in \cite{SinghSidhu:2020cxw}. In this case a human's  death may occur as a result of  a large dose of radiation with a long time delay, which would make  it hard to identify the cause of the death. This argument (about the time delay and difficulties with possible death identification) should be contrasted   with the main assumption of \cite{SinghSidhu:2020cxw} that all such cases would be unambiguously and quickly identified.}.
}
 While ground based direct searches   
offer the most unambiguous channel
for the detection of quark nuggets   
the flux of nuggets   is inversely proportional to the nugget's mass   and 
consequently even the largest available conventional dark matter detectors are incapable  to exclude    the entire potential mass range of the nuggets. Instead, the large area detectors which are normally designed for analysing     the high energy CR are much better suited for our studies of the AQNs as we discuss in next section \ref{earth}.

 \subsection{When the AQNs hit the Earth...}\label{earth}
    For our present work, however,  the  most relevant studies  are related to the effects which  may occur when the AQNs made of antimatter hit the Earth  and continue to  propagate in deep underground in very dense environment.
   In this case the  most of the energy deposition will   occur  in the Earth's interior.    The corresponding signals are very hard to detect  as the photons, electrons and positrons will be quickly absorbed by surrounding dense  material deep underground, while the emissions of the very weakly interacting neutrinos and axions are hard to recover. 
   \exclude{
   Nevertheless, as we already mentioned,   the AQN-induced neutrinos produced in the Earth's interior might be responsible  for explanation of the long standing puzzle of the DAMA/LIBRA  observation  of the annual modulation  \cite{Zhitnitsky:2019tbh}. The AQN-induced axions from deep interior can be recovered by analyzing    the daily and annual modulations  as suggested in \cite{Budker:2019zka} and elaborated in \cite{Liang:2020mnz}.      The AQN annihilation events in the Earth's atmosphere could produce  infrasound and seismic acoustic waves     as discussed in   \cite{Budker:2020mqk,Figueroa:2021bab}  when the infrasound and seismic acoustic waves indeed have been recorded  by  dedicated instruments\footnote{A single observed event properly recorded by  the Elginfield Infrasound Array  (ELFO) which was  accompanied by correlated seismic waves is dramatically different from conventional meteor-like  events. In particular, while the event was very intense it has not been detected by a synchronized  all-sky camera network (visible frequency bands)  which ruled out a meteor source. At the same time this event is consistent  with   interpretation of the    AQN-induced    event because  the visible frequency bands must be strongly suppressed when  AQN propagates in atmosphere \cite{Budker:2020mqk} }. 
   }
    In this short subsection we want to mention several   observed phenomena which could be related to   the AQN annihilation events when the nuggets propagate in the Earth's atmosphere. 
    
    We start with the Telescope Array (TA) experiment which has recorded  \cite{Abbasi:2017rvx,Okuda_2019}  several short time bursts of air shower like events. 
       These events are very unusual and cannot be interpreted  in 
terms of  conventional  CR single showers. In particular, if one tries to fit the observed bursts (cluster events) with conventional code, the energy for CR events should be in $10^{13}$ eV energy range to match the frequency of appearance, while the observed bursts correspond to $(10^{18}-10^{19})$ eV energy range as estimated by signal amplitude and distribution. Therefore, the estimated energy from individual events within the bursts is five to six orders of magnitude higher than the energy estimated by event rate  \cite{Abbasi:2017rvx,Okuda_2019}.    It has been argued in  \cite{Zhitnitsky:2020shd,Liang:2021wjx}    that these bursts  represent  the direct manifestation of the
   AQN   annihilation events.

This feature,  in many aspects,  is very similar to the {\it intensity puzzle} listed in previous Section \ref{confronts}, where analogous  dramatic inconsistency (between the observed intensity and measured event rate) is recorded  by H10T instrument.

Our next example  is  the observed seasonal variations of the X ray   background in the near-Earth environment. To be more precise,   the  XMM-Newton at  $ 11\sigma$ confidence level  \citep{Fraser:2014wja} had recorded the seasonal variations in the 2-6 keV energy range  at distances $r\gtrsim 8 R_{\oplus}$ where the measurements are effectively had been performed. The authors \citep{Fraser:2014wja} argue that conventional astrophysical sources have been ruled out. Furthermore, the XMM-Newton's operations exclude pointing at the Sun and at the Earth directly, diminishing a possible direct X ray  background from the Earth and the Sun. It has been argued in   \cite{Ge:2020cho} that this seasonal variation may be  naturally explained  by the AQNs exiting the Earth.  The AQNs   continue the emission of   the X rays at  $r\gtrsim 8 R_{\oplus}$,  long after the nuggets  exit the Earth's atmosphere. It has been also shown that the spectrum and the intensity computed in  \cite{Ge:2020cho}  almost identically match the observed spectrum \citep{Fraser:2014wja}. 
 
 The final example we want to mention here is the observation by ANITA \cite{Gorham:2016zah,Gorham:2018ydl,ANITA:2020gmv} of   two anomalous events with noninverted polarity. These two events had been  identified as the Earth emergent upward going CR-like events    with exit angles  of $ -27\degree$ and $ -35\degree$   relative to the horizon. These anomalous events  are in dramatic tension with the standard model   because neutrinos are exceedingly unlikely to traverse through Earth at a distance of $\gtrsim5\times10^3\,$km with such ultrahigh energy, even accounting for the $\nu_\tau$ regeneration \cite{Gorham:2016zah}. The analysis \cite{Fox:2018syq} reviewed the high-energy neutrino events from the IceCube Neutrino Observatory and inferred  that the $\nu_\tau$ interpretation is excluded by at least $5\sigma$ confidence level.    We  advocated in \cite{Liang:2021rnv} that  these two anomalous events with noninverted polarity observed by ANITA could be interpreted as the upward going AQNs.   

 The  applications  \cite{Ge:2020cho} to the X ray emission in the near-Earth environment  and the ANITA anomalous events \cite{Liang:2021rnv}   are  especially relevant for the present work because both these phenomena     are  originated from the AQN upward going 
 (Earth emergent) events when the AQNs traversed  through the Earth interior and exit the Earth surface.  
 As we shall argue in this work the MME events could be also the consequence of the same  upward moving AQNs.   
 
 However, there is an additional technical challenging problem to  be addressed for the application to MME analysis. 
 The point is that the ANITA anomalous events are generated by AQNs which just crossed the surface and entered the Earth's atmosphere (upward-going events). Exactly at this instant large number of electrons instantaneously released into atmosphere and 
 generated the radio pulse measured by ANITA. In contrast, the X rays are observed   by XMM-Newton at very large distances $r\gtrsim 8 R_{\oplus}$ from Earth. In this case   the   impact of the atmospheric material at $r\lesssim 60$ km can be ignored, and the cooling of the AQN is dominated by nugget's propagation  in an empty space.  The application to MME, which is the topic of the present work,  deals with an intermediate stage between these two cases: it is not the first  instant 
 when the AQN enters the atmosphere emerging from the interior, and it is not the asymptotically far away  when the AQN propagates in empty space. 
 In other words, we need to know the emission pattern and the energy deposition rate for  the AQNs when they propagate in the atmosphere, and the impact of the atmospheric surrounding material cannot be ignored. We formulate our proposal in Section \ref{proposal} where we argue that 
  emission of the electrons  by the AQN   in form of the well-separated ``bunches"   may explain the unusual features of the MME observations.  
 
 However, before we present our arguments supporting this interpretation of the MME unusual events     (as  highlighted  in section \ref{confronts})  we should overview  the basic characteristics of the AQNs traversing the Earth, which represents  the topic of the next subsection  \ref{internal}.   
    
     \subsection{Internal structure of the AQN}\label{internal}
     
      The goal  here is to explain the basic features of the AQNs when they enter the dense regions of the surrounding material and annihilation processes start. 
       The related  computations originally have been carried out in \cite{Forbes:2008uf}
 in application to the galactic environment with a typical density of surrounding visible baryons of order $n_{\rm galaxy}\sim 300 ~{\rm cm^{-3}}$ in the galactic center, in dramatic contrast with dense region in the Earth's interior  when $n_{\rm rock}\sim 10^{24} ~{\rm cm^{-3}}$ and atmosphere with  $n_{\rm air}\sim 10^{21} ~{\rm cm^{-3}}$. We review  these computations with few additional elements which must be implemented in case of propagation in the  Earth's atmosphere  and interior when the density of the environment is much greater than in the galactic  environment.  

The total surface
emissivity from electrosphere has been computed in \cite{Forbes:2008uf} and it is given by 
\begin{equation}
  \label{eq:P_t}
  F_{\text{tot}} \approx 
  \frac{16}{3}
  \frac{T^4\alpha^{5/2}}{\pi}\sqrt[4]{\frac{T}{m}}\,,
\end{equation}
 where $\alpha\approx1/137$ is the fine structure constant, $m=511{\rm\,keV}$ is the mass of electron, and $T$ is the internal temperature of the AQN.  
 One should emphasize that the emission from the electrosphere is not thermal, and the spectrum is dramatically different from black body radiation, see \cite{Forbes:2008uf},  see also Appendix \ref{sect:kappa} with more details. 
   
A typical internal temperature   of  the  AQNs for very dilute galactic environment can be estimated from the condition that
 the radiative output of Eq. (\ref{eq:P_t}) must balance the flux of energy onto the 
nugget 
\be
\label{eq:rad_balance}
    F_{\text{tot}}  (4\pi R^2)
\approx \kappa\cdot  (\pi R^2) \cdot (2~ {\rm GeV})\cdot n \cdot v_{\rm AQN},  
\ee 
where $n$ represents the number density of the environment.  The left hand side accounts for the total energy radiation from the  AQN's surface per unit time as given by (\ref{eq:P_t})  while  
 the right hand side  accounts for the rate of annihilation events when each successful annihilation event of a single baryon charge produces $\sim 2m_pc^2\approx 2~{\rm GeV}$ energy. In Eq.\,(\ref{eq:rad_balance}) we assume that  the nugget is characterized by the geometrical cross section $\pi R^2$ when it propagates 
in environment with local density $n$ with velocity $v_{\rm AQN}\sim 10^{-3}c$.

The factor $\kappa$ is introduced to account for the fact that not all matter striking the  AQN will 
annihilate and not all of the energy released by an annihilation will be thermalized in the  AQNs by changing the internal temperature $T$. In particular,  some portion of the energy will be released in form of the axions, neutrinos\footnote{\label{kappa}In a neutral dilute galactic environment considered  previously \cite{Forbes:2008uf}  the value of $\kappa$ was estimated as  $\kappa\approx 0.1$.}. The high probability 
of reflection at the sharp quark matter surface lowers the value of $\kappa$. The propagation of an ionized (negatively charged) nugget in a  highly ionized plasma (such as solar corona)   will 
 increase  the effective cross section. As a consequence,   the value of  $\kappa$ could be very large as discussed in \cite{Raza:2018gpb} in application to the solar corona heating problem.
 
 The  internal AQN temperature  had been estimated previously for a number of  cases.  It  may assume dramatically different values, mostly due to the huge difference in number density $n$ entering (\ref{eq:rad_balance}). In particular, for the galactic environment $T_{\rm galaxy}\approx 1$\,eV, while in 
 deep Earth's interior it could be as high as  $T_{\rm rock}\approx  (100-200) $\,keV. Precisely this value of  $T$  had been used as initial temperature 
of the nuggets in  the proposal \cite{Ge:2020cho} explaining the seasonal variations of the X rays observed the  XMM-Newton at  $ 11\sigma$ confidence level  \citep{Fraser:2014wja} at distances $r\sim (6-10) R_{\oplus}$ from the Earth surface. 

The crucial element for the study \cite{Ge:2020cho}  was the   emission  rate by AQNs at very high temperatures $T \approx  (100-200) $\,keV. In this case the emission   is not determined by simple Bremsstrahlung radiation given by (\ref{eq:P_t}) which is only valid   for relatively low temperatures. In the high  temperature regime a number of many-body effects  in the electrosphere, that were previously ignored, become important. In particular, it includes: the generation of the plasma frequency in electrosphere, which suppresses the low frequency emission. Also, the the AQNs get ionized  which   dramatically decreases the density of positrons in electrosphere, and consequently suppresses the emission, among many others effects. All these complications   have been carefully considered in  \cite{Ge:2020cho}  in  applications to  the  seasonal variations of the X rays as observed the  XMM-Newton, and we refer to that paper for the details. Here we quote  the result of these studies. The rate of emission can be effectively represented as follows:
\be
\label{cooling}
\frac{dE_{\rm emiss}}{dt}  \sim (4\pi R^2)   \eta(T,R) F_{\text{tot}} (T)
\ee  
where   factor  $\eta(T,R)$ is a result of strong ionization of the electrosphere, which leads to  the corresponding suppression of the emission. The details of this suppression are explained in Appendix \ref{sect:kappa}
and given  by    (\ref{ratio1}). This suppression  is a direct consequence of high internal temperature $T$ when a large number of  weakly bound positrons     are expanded over much larger distances order of $R$ rather  than distributed over  much shorter distances of order $m^{-1}$ around the nugget's core. This basically determines  the   suppression factor $\eta\sim (mR)^{-1}\sim 10^{-6}$.

    In both previously considered  applications (to ANITA \cite{Liang:2021rnv} and to XMM-Newton  \cite{Ge:2020cho} experiments) when the upward moving AQNs with  high $T \approx  (100-200) $\,keV play the key role in the explanations of the observations, we ignored an additional\footnote{\label{additional}We use the term``additional" to emphasize that the AQN had accumulated a huge amount of energy during the transpassing   the Earth interior  as the capacity of the quark core nugget is very large \cite{Ge:2020cho}. Precisely this accumulated energy will be emitted in form of the X rays when the AQN propagates in empty space at distances $r\sim (6-10) R_{\oplus}$ from the Earth surface.} annihilation events with atmospheric material as we already mentioned at the end of the previous subsection \ref{earth}.  Now we want to include the corresponding physics in our analysis.

First of all we want to estimate the number of direct head-on collisions of the atmospheric molecules with AQN per unit time. It  can be estimated as follows:
\be
\label{collisions}
\frac{dN}{dt}\simeq (\pi R^2)    N_{\rm m}  v_{\rm AQN}
\simeq 0.8\cdot 10^{18} \left(\frac{n_{\rm air}}{10^{21} ~{\rm cm^{-3}}}\right)\rm s^{-1},~ ~~
\ee   
where $N_m\simeq 2.7\cdot 10^{19}  ~{\rm cm^{-3}}$ is the molecular density  in atmosphere when each molecule contains approximately 30 baryons such that the typical density of surrounding baryons in air is  $n_{\rm air}\simeq 30\cdot N_m\simeq 10^{21} ~{\rm cm^{-3}}$.
The dominant portion  of these collisions are the elastic scattering processes rather than successful annihilation events  suppressed by parameters $\kappa$ as discussed above. The energy being deposited to the AQN per unit time as a result of annihilating processes can be estimated as follows   
 \be
\label{energy}
&&\frac{dE_{\rm deposit}}{dt} \approx   (2~ {\rm GeV})\cdot \kappa\cdot \frac{dN}{dt} \\
&\approx& 1.6\cdot 10^{18} \cdot \kappa\cdot  \left(\frac{n_{\rm air}}{10^{21} ~{\rm cm^{-3}}}\right)\rm\frac{GeV}{s},\nonumber
\ee   
One should emphasize that this  energy is being  deposited into the AQN which is already at full capacity with very high   temperature $T\simeq (100-200)$ keV  accumulated during a long journey in much  denser Earth's interior.  

What is the mechanism to release this extra energy? In other words: In what form  the energy (\ref{energy}) could be emitted  into the surrounding atmosphere  as it cannot be easily transferred to the   AQN's quark core (as it is already saturated, see footnote \ref{additional})?  

There are two qualitatively different   regimes  which can be normally   realized in such circumstances. If a typical time scale $\tau_{deposit}$ to deposit a specific amount of energy (let us say, 1 GeV)      determined by (\ref{energy}) is   longer  than the time scale $\tau_{\rm cool}$ to release the same amount of energy, i.e. $\tau_{deposit} \gtrsim \tau_{\rm cool}$ than a continuous cooling process takes place and the temperature slowly decreases. This is  a thermodynamically equilibrium behaviour  which can be analyzed using conventional technical tools\footnote{\label{ft:cooling}For example,  the AQNs moving in empty space and slowly cooling by emitting the X rays as computed in \cite{Ge:2020cho}  belongs to this class.}.

The second option is realized  when the time scale to deposit the  energy $\tau_{deposit}$  is much  shorter than the  time scale of     cooling processes, i.e. $ \tau_{deposit}\ll \tau_{\rm cool}$, 
in which case the burst-like (explosion like, eruption-like blasts)  non- equilibrium  processes must occur. These eruption-like events will release the extra accumulated energy in form of very short  pulses, which cannot be described as conventional cooling  equilibrium processes.   These short pulses must  alternate with much longer periods of accumulation  energy which inevitably end with subsequent eruption-like events.  In other words, in this case, the non-equilibrium fast equilibration process can be thought as  the 
  eruption-like event\footnote{\label{analogy1}An analogy for such eruption-like event is the lightning flash  under thunderclouds. The clouds accumulate the electric charge  in form of the ionized molecule  very efficiently. The corresponding time scales    plays the role of  $\tau_{deposit}$ in  our system. The neutralization of these ions is less efficient process, which is analogous  to our    $\tau_{\rm cool}$. If some conditions are met (the so-called runaway breakdown conditions are satisfied, see \cite{Gurevich_2001, DWYER2014147} for review) the discharge occurs in form of the eruption which is the lightning strike in our analogy. The system is getting neutralized in form of the non- equilibrium   lightning event (eruption) on the time scales which are much shorter than any time scales of the problem. This analogy is in fact, quite deep, and will be used in   section \ref{timing} when we discuss different time scales of MMEs as observed by H10T detector.}.   

   Now we   give a numerical estimation for  the cooling rate   determined by (\ref{cooling}) to  find out  what regime  is realized in given circumstances.
The order of magnitude estimate can be  expressed  as follows:
 \be
\label{cooling1}
\frac{dE_{\rm emiss}}{dt}  \sim 10^{16}\left(\frac{T}{100~ \rm keV}\right)^{\frac{17}{4}} \rm\frac{GeV}{s}.
\ee  
Comparing this estimate for cooling rate with estimate for the deposition rate (\ref{energy}) one concludes that the rate  of energy deposition 
     is much shorter    than the rate at which the system is capable to release the same amount of energy  in form of the photon's 
     emission. As a result, we arrive to the conclusion  that second option 
when  $ \tau_{deposit}\ll \tau_{\rm cool}$  is realized.
This conclusion    inevitably implies the eruption like events  must occur, and  these very short explosions  must be    alternated with much longer periods of time when the  energy is being accumulated by the AQNs. These   longer periods  of the energy accumulations also must  end with consequent  eruption-like blasts. This alternating pattern continues as long as following condition holds:

\be
\label{condition}
  \tau_{deposit}\ll \tau_{\rm cool}  ~~\Rightarrow~~{\rm  (short~ eruptions~ occur)}. 
\ee
With these preliminary comments  with estimates from the previous studies  on the AQN features we  are  now in position to formulate the proposal  interpreting  the   MME events observed by X10T collaboration in terms of  the   upward moving   AQNs, which is the topic of the next section \ref{proposal}.

\section{MME  as the  AQN      event}\label{proposal}
 In this Section  we formulate  the basic idea of our proposal on identification of the unusual Multi-Modal Events with the  upward moving AQN events, while supporting estimates on the event rate, the intensity, and the variety of time scales will be presented in following sections  \ref{sec:rate} and \ref{timing}.

The AQNs which propagate in the Earth's interior are very hot. Their temperature could be as hot as $T\approx (100-200)$\,keV at the moment of exit on the Earth's surface as we discussed in 
 \cite{Ge:2020cho} in the application to studies of the seasonal variations of the X rays observed   by XMM-Newton at very large  
 distances $r\sim (6-10) R_{\oplus}$ from the Earth.  At such large distances the AQN's cooling can be described by conventional thermodynamically equilibrium processes   with well defined spectrum,  see footnote \ref{cooling}.  
 It should be contrasted with our application to ANITA  anomalous events \cite{Liang:2021rnv} when    the same upward moving 
 AQNs  just crossed the surface and entered the Earth's atmosphere.  In this case the dominant  cooling mechanism 
 is realized in form of short burst-like events according to (\ref{condition}). The outcome of these short burst-like events is emission of numerous  bunches (clumps) of relativistic electrons.  These eruption like events are analogous to lightning flashes  from footnote \ref{analogy1}.
 
 In our studies   \cite{Liang:2021rnv}  we focused on an estimation of 
 the number  of emitted electrons $N$ and their energy $\la E\ra \sim 10~ {\rm  MeV}$  which, according to the proposal  \cite{Liang:2021rnv}, essentially determine the intensity and the spectral features of the ANITA  anomalous events. 
 These electrons will be always accompanied by the emission of much more numerous number of  photons with typical energy of order $\omega\sim T\approx (100-200)$\,keV, which in fact represent the dominant fast non- equilibrium mechanism of cooling  of the system in  these circumstances  (\ref{condition}). 
 
 Precisely these multiple  explosion-like emissions   of photons and electrons will be identified with  Multi-Modal Events  with their unusual features  reviewed in Section \ref{confronts}.
 The mean free path of the photons with $\omega\sim T$ is relatively short, measured in meters.  Therefore, they cannot be directly observed.  In contrast, 
 the mean free path of the  energetic electrons at sea  level with $\la E\ra \sim 10~ {\rm  MeV}$  is around 
  several kilometres in atmosphere (and even  longer at the elevation of 3346 m where H10T is located). The propagation of these ultra-relativistic electrons  takes place  in the background of the geomagnetic field ${\cal{B}}\sim 0.5 ~\rm G$ which determines  the instantaneous  curvature 
 $\rho\sim   3$ km of the electron's trajectories with such energies, see section \ref{sec:rate} with relevant estimates. Due to the large component of magnetic field parallel to the Earth's surface the electrons may move in downward direction even if they are  initially emitted  in  upward direction.    Our proposal is that precisely these downward moving energetic electrons  
   could mimic the CR and could be   responsible for the Multi-Modal pulses  as   detected by H10T instrument.

Any precise computation  during  a  short period of time  of emission   is very hard problem  of non-equilibrium dynamics, which is beyond the scope of the present work. Fortunately, the observable intensity, the event rate, the   time delays between the pulses    are determined in the AQN framework  by the basic characteristics of the   model 
 and are  not very sensitive to the details of this non-equilibrium mechanism of production and its corresponding time scales. This is because all these observables   depend on several   parameters such as typical energy $\la E\ra \sim 10~ {\rm  MeV}$ of the emitted electrons,  typical number of emitted electrons $N$,
 the deposition rate (\ref{energy}) and the cooling rate (\ref{cooling1}) which had been previously estimated for very different purposes in different circumstances. We shall use the same parameters in   the present  work.

 In particular, the number of electrons $N\sim (10^8-10^{9})$ emitted by AQNs  is very hard to compute 
 from the first principles. However it can be  fixed  by the  observed intensity and spectral features of the  anomalous ANITA radio pulses
 assuming, of course, that these observed  radio pulses with inverted polarities are due to the upward moving AQNs  \cite{Liang:2021rnv}. 
 The  typical  energy $\la E\ra \sim 10~ {\rm  MeV}$ of these electrons is also consistent with duration of the observed    pulses \cite{Liang:2021rnv}.  The next section \ref{sec:rate} is devoted to the estimation of the  event rate  of   MME, while section \ref{timing}  is devoted to explanations of the {\it puzzles} which demonstrate   a dramatic deviation from  conventional CR picture as formulated in section \ref{confronts}. We shall argue that the observed MME features are   consistent  with our AQN-based interpretation.   
    
\section{Event Rate of MME }
\label{sec:rate}
 This section is devoted to estimate of the event rate of MMEs within the AQN framework. We anticipate that  this evaluation  is expected to  be very qualitative estimation    due to large uncertainties in parameters and rare occurrence of the observed MMEs. In particular, it is known that key parameter, the   DM density locally  may dramatically deviate  from  the well established average global value $\rho_{\rm DM}\approx 0.3{\rm\,GeV\,cm^{-3}}$. Nevertheless we would like to present such estimate to demonstrate that our interpretation of MMEs as an outcome  of upward-going AQNs is at least a self-consistent proposal. In what follows we use the same formulae we used for the estimates of the frequency of appearance of the  ANITA  Anomalous  Events \cite{Liang:2021rnv} such that the  numerous uncertainties related to the local DM density or/and the AQN size distribution 
 would not dramatically affect our estimates below if they are normalized to the ANITA event rate.
 
 We start  with the same formula from  \cite{Liang:2021rnv}  for the expected number ${\cal N}$ of the MMEs  assuming that they are induced by the AQNs:
\begin{equation}
\label{eq:cal N}
{\cal N}
\approx {\cal{A}}_{\rm eff} {\cal T}\Delta\Omega
\frac{\rmd\Phi}{\rmd A\rmd\Omega}\,,
\end{equation}
where  $\Delta\Omega\approx 2\pi$ for the isotropic AQN flux, and the expression for the local rate of upward-going AQNs per unit area is given by:
\be
\label{Phi1}
&& \frac{\rmd\Phi}{\rmd A\rmd\Omega}
\approx\frac{\Phi}{4\pi R_\oplus^2}  \approx  4\cdot 10^{-2}\left(\frac{10^{25}}{\langle B\rangle}\right)\rm \frac{ events}{yr\cdot  km^2},\\
&&\Phi
\approx\frac{2\cdot10^7}{\rm yr  }
 \left(\frac{\rho_{\rm DM}}{0.3{\rm\,GeV\,cm^{-3}}}\right)
\left(\frac{v_{\rm AQN}}{220\kmps}\right)
\left(\frac{10^{25}}{\langle B\rangle}\right), \nonumber
\ee
where $\rho_{\rm DM}$ is the local density of DM and  
 $R_\oplus=6371\,$km is the radius of the Earth and $\Phi$ is the total hit rate of AQNs on Earth \cite{Lawson:2019cvy}.
 In formula (\ref{eq:cal N})    the
 ${\cal T} $ is the time  of operation    while  ${\cal{A}}_{\rm eff} $ is the effective area to be estimated below. 
 
 The estimation of the effective area ${\cal{A}}_{\rm eff} $ is a complicated task as it is not simply determined by the area of the detectors similar to standard CR analysis. Instead, it is determined by the area along the AQN's path where  it   can emit the bunches of electrons in form of short pulses which can mimic the CR as outlined in previous section \ref{proposal}. The area ${\cal{A}}_{\rm eff} \sim L_{\rm AQN}\cdot D_{\rm AQN}$ can be thought as a strip of length 
 $L_{\rm AQN}$ and width $D_{\rm AQN}$.  
 
To estimate these parameters  let us consider an  AQN   moving in upward direction with angle $\theta_{\rm AQN}$ with respect  to the Earth's surface
 such that upward velocity component is $(v_{\rm AQN}\cdot \sin \theta_{\rm AQN})$. This vertical velocity component determines the maximal height $h$ of the atmosphere where the atmospheric density is still sufficiently high such that condition (\ref{condition})   holds. We estimate  $h\approx (30-50)$ km where the atmospheric density falls by two order of magnitude
 such that conventional  X-ray emission (which had been used  in computations \cite{Ge:2020cho} to explain the seasonal variations  observed by XMM Newton)   becomes the dominant cooling process at higher altitudes. 
 This condition  determines the maximal length $L_{\rm AQN}$  and time  scale $\tau_{\rm AQN}$ when the AQNs can emit the bunches of electrons in form of pulses which we identify with MMEs and which can mimic the CR air showers as outlined in previous section \ref{proposal}. Numerically these parameters are estimated as follows: 
 \be
 \label{parameters}
 \tau_{\rm AQN}&\sim& \frac{h}{(v_{\rm AQN}\cdot \sin \theta_{\rm AQN})}\sim 0.2 \rm s,\\
 L_{\rm AQN}&\sim& h\cot\theta_{\rm AQN}\sim (50-90) \rm km, \nonumber
 \ee
  where we used $\theta_{\rm AQN}\simeq 30^0$ for an order of magnitude numerical estimates.
  
  To estimate the width of the strip $D_{\rm AQN}$ we  recall that the geomagnetic field ${\cal B}$  in  location  of the H10T instrument  can be  characterized by two numbers: it has strong component in up to down direction ${\cal B}_{\rm down}\simeq 0.49$ G, and strong component from south to north  direction ${\cal B}_{\rm north}\simeq 0.24$ G, see e.g. \cite{geomagnetic}.
  
  Assuming that the energetic electrons with $\la E\ra \sim 10~ {\rm  MeV}$ will be emitted along the   ${\mathbf{v}}_{\rm AQN}$-direction one can estimate  instantaneous radius of curvature $\rho$, see  e.g. Jackson \cite{jackson}:
\begin{equation}
\label{eq:rho}
\rho
\approx\frac{\gamma mc}{e{\cal B}  \sin\theta_{\cal B}}
\approx 2.8{\rm\,km}
\left(\frac{\gamma}{20}\right), ~~~~~~ \gamma\equiv \frac{\la E\ra}{m}
 \end{equation}
where ${\cal B}\approx {\cal B}_{\rm north}\approx 0.24$ G is the local magnetic field strength, which changes the direction of the electrons from upward   to downward moving such that they can mimic the CR air showers. The angle  $\theta_{\cal B}$ is the angle between the particle electron velocity $\mathbf{v}$ and magnetic direction. We choose $\theta_{\cal B}\approx 30\degree$ in (\ref{eq:rho})  for the numerical estimates. 

 Now we estimate the relevant scale  $\lambda$ which determines the survival pattern of the electron's bunches.
 It  is mostly determined by the Coulomb elastic scattering with cross section $\sigma_{\rm Coul}$  
     \be
    \label{Coulomb}
    \sigma_{\rm Coul}\approx \frac{\alpha^2}{E^2\theta^4} \approx 3\cdot 10^{-27} \left(\frac{1/2}{\theta}\right)^4 \left(\frac{20}{\gamma}\right)^2\rm cm^2 .
        \ee
     The electrons with $\theta\gtrsim 1/2$ may strongly deviate from  their main paths and cease to stay with majority of particles forming the bunch (which eventually becomes the pulse  being interpreted as MME).
   The corresponding length scale $\lambda(h)$ at altitude $h$ (accounting for the air-density $n_{\rm air}(h)$ variation) is estimated as follows
     \be
   \label{lambda1}
   \lambda (h) \sim    \frac{1}{ \sigma_{\rm Coul}  n_{\rm air}(h)}\sim 3\cdot \exp\left({\frac{h}{8~ \rm km}}\right) ~\rm km   .
   \ee
   The physical meaning of $\lambda$ is the length distance particles propagate  at which the majority of the particles in  the bunch remain within the bunch before the dispersing to much larger distances when they cease to be a part of the pulse (burst). 
Important  comment  here is that $\lambda\gtrsim \rho$ such that the majority of electrons within  the bunch do not strongly re-scatter,  and therefore survive the    change in orientation due to the geomagnetic field ${\cal B}$  (from upward to downward direction).  

The width of the strip $D_{\rm AQN}$ now can be roughly estimated as follows:
\be
 \label{parameters1}
  D_{\rm AQN}\sim   \lambda\sim  3\cdot \exp\left({\frac{h}{8~ \rm km}}\right) \cdot  \left(\frac{\gamma}{20}\right) {\rm\,km}. 
 \ee
Combining all the estimates above one arrives to the following order of magnitude estimation for the expected number of MME events:
\begin{equation}
\label{eq:cal N ::num}
{\cal N}
\approx0.5\cdot 10^2 \rm events
\left(\frac{{\cal{A}}_{\rm eff}}{500 ~\rm km^2}\right)\left(\frac{{\cal{T}} }{0.4 ~\rm yr}\right). 
 \end{equation}
 This order of magnitude estimate should be compared with observed frequency of appearance of the MMEs. According to \cite{2017EPJWC.14514001B}   the observations from start until August 2016 during ${\cal T}\approx \,3,500 \rm h \approx 0.4 \rm yr$   of operation  the H10T instrument  had detected ${\cal N}_{\rm obs}\simeq 10^3$  MMEs. According to  \cite{Beznosko:2019cI}   the observations from Feb 15, 2018 to May 12,  2018 (which corresponds to ${\cal{T}} \approx {0.25 ~\rm yr}$)  the H10T instrument  had detected ${\cal N}_{\rm obs}= 217$  MMEs.
 
The AQN based estimations   (\ref{eq:cal N ::num}) suggest that  number of events ${\cal N}$ is    almost one order of magnitude lower than  ${\cal N}_{\rm obs}$ events  observed by H10T. Nevertheless, we consider this order of magnitude estimation  (\ref{eq:cal N ::num}) being  consistent with our proposal due to many uncertainties which enter this estimate. There are several reasons for this optimistic view.

First, as we already mentioned the parameters entering    \eqref{Phi1} are in fact not precisely known. The fundamental  parameters such as $\rho_{\rm DM}$ and $\langle v_{\rm AQN}\rangle$  only have accuracy up to order one as the local flux distribution of DM and size distribution of AQN remain unknown to date as the   DM density locally  may dramatically deviate  from  the well established average global value $\rho_{\rm DM}\approx 0.3{\rm\,GeV\,cm^{-3}}$, see introduction in \cite{Liang:2020mnz} for the references and details.    Furthermore, the size distribution factor entering  \eqref{Phi1} in form $\la B\ra^{-1}$ had been fixed from dramatically different physics (including solar corona heating puzzle) and  can easily deviate by large factor. The estimation of the effective area ${\cal{A}}_{\rm eff} $ also suffers enormous uncertainty.

 Furthermore,     the  numerical   suppression factor $\sim 0.1$ (between computed ${\cal N}$ and observed ${\cal N}_{\rm obs}$ values)  which appears in our  order of magnitude estimates  (\ref{eq:cal N ::num}) is very similar to    suppression  factor $\sim 0.1$ which occurred in analogous  estimates   for the mysterious TA bursts \cite{Zhitnitsky:2020shd}  and the ANITA anomalous events   \cite{Liang:2021rnv}.  This similarity hints on a common   origin for all these  phenomena.   Therefore, if one normalize ${\cal N}$ to the observed ANITA anomalous events (assuming that all three  phenomena originated from the same AQN based physics) one arrives to correct order of magnitude estimate as the suppression factor $\sim 0.1$ is approximately the same for all three  phenomena  estimated for the mysterious  TA bursts in \cite{Zhitnitsky:2020shd},   the ANITA anomalous events  in  \cite{Liang:2021rnv}, and Multi-Modal Events estimated in (\ref{eq:cal N ::num}).  
 
 Our main arguments, however,  leading to the  identification of the MMEs with the AQN events 
are  not based on an order of magnitude estimate of the frequency of appearance   (\ref{eq:cal N ::num}) which is presented in this work exclusively for illustrative purposes. 
 Rather, our main arguments are based on specific qualitative features (formulated as {\it puzzles}) which have been listed  in Section \ref{sec:introduction}, and which cannot be 
explained in terms of the conventional CR air showers as highlighted in Section \ref{confronts}.    We consider a qualitative agreement   between the observations and our theoretical estimates  (to be discussed in next section \ref{timing})   as the  strong  arguments supporting our identification.

 \section{AQN   proposal  confronts  the   MME observations \cite{Beisembaev:2016cyg,2017EPJWC.14514001B,Beznosko:2019cI,2019EPJWC.20806002B,Beisembaev:2019nzd}} \label{timing}
 We start this section by estimating  the  particle number density $\rho(R)$  which appears in formulation of the {\it puzzles} in Section \ref{confronts}.
   The particle number density in the AQN framework is determined by the number of particles which 
  could be observed at the detector site.  Assuming that the total number of electrons being emitted in form of a pulse (as a result of erupted burst) is similar to the number of particles which was used in our analysis \cite{Liang:2021rnv} of the ANITA anomalous events $N\simeq (10^8-10^9)$    
  we arrive to the following estimate for   $\rho(R)$:
 \be
 \label{rho}
 \rho(R)\sim \frac{N}{(\lambda\Delta\theta)^2} \sim \frac{(10^8-10^9)}{(2.5~ \rm km)^2}\sim  \frac{(20-200)}{\rm m^2}
 \ee
 where    the area which is hit by the bunch of particles from  a single  burst is estimated as $(\lambda\Delta\theta)^2$ with  $\lambda(h)$ given by  
  (\ref{lambda1}) at the detector site with $h\simeq 4 $ km. It is assumed that  the  particles  are propagating within the cone  with angle $\Delta\theta\lesssim 1/2$.
  
  Our next task is the estimation of the time delay between the pulses within the same cluster. The corresponding time delay can be estimated from the condition that 
  the energy deposited into the AQN during time $\tau_{\rm delay}$ must be released  in form  of the eruption when $N\simeq (10^8-10^9)$     electrons being emitted in form of a single pulse. As we mentioned previously, the dominant portion of the energy in such eruption is released in form of the photons with energy $\omega\sim T$ which always accompany the   electron's emission as a result of these bursts. The relative ratio $r$ between these two components is estimated in Appendix \ref{pair-production}, see eq. (\ref{ratio-Appendix}).  Numerically, the estimate can be presented as  follows:
     \be
  \label{ratio}
 r\equiv \frac{\la E_{e^+e^-}\ra}{\la E_{\gamma}\ra}\sim \frac{\alpha}{2\pi}\exp \left(-\frac{2m}{T}\right)\sim 0.7\cdot 10^{-5},  
   \ee
   where for numerical estimates we use $T\simeq 200$ keV.
  We must emphasize that  (\ref{ratio})  is really an order of magnitude estimate as mechanism of emission is determined by a very complex physics as highlighted in   Appendix \ref{pair-production}. Furthermore, the corresponding eruption event  is not a thermodynamically equilibrium process as we  already  mentioned previously.  
  The dominant contribution  in form of the photon's emission during  a single pulse can be estimated  in terms of these parameters as follows
  \be
  \label{eq:pulse}
  \la E_{\gamma}\ra\sim N \cdot (2m)\cdot r^{-1} \sim 1.4 \cdot (10^{10}-10^{11})~  \rm GeV,~~~~~~
   \ee
   where we use the numerical value for   $N\in (10^8-10^9)$ extracted from   analysis of the ANITA anomalous events  \cite{Liang:2021rnv} and parameter $r$ is given by (\ref{ratio}). In this estimate we assumed that the energy of the pair $\la E_{e^+e^-}\ra \sim 2mN$  computed at the moment of the pair creation.  
  The  released energy during a single pulse (\ref{eq:pulse}) determines the time delay $\tau_{\rm delay}$ between the pulses within the same cluster:
  \be
  \label{eq:delay}
  \tau^{-1}_{\rm delay}\sim \frac{ {dE_{\rm deposit}}/{dt} }{\la E_{\gamma}\ra}\sim \kappa \frac{ 10^{18} \cdot  \rm {GeV} s^{-1}}{ (10^{10}-10^{11})~  \rm GeV},
  \ee
  where we used numerical value for $ ({dE_{\rm deposit}}/{dt}) $ from (\ref{energy}) with density of atmosphere at $h\simeq 4$ km where the 
  detector H10T is located.   Our final, the   order of magnitude estimate for the time delay $\tau_{\rm delay}$ between the pulses can be represented as follows
  \be
  \label{eq:delay1}
  \tau_{\rm delay}\sim \left(\frac{0.1}{\kappa}\right)\cdot \left(10^2-10^3\right) \rm ns,
    \ee
    where we use $\kappa\simeq 0.1$ in (\ref{eq:delay1}) for numerical estimates which was previously extracted from studies  of the emission in the dilute galactic environment, see footnote \ref{kappa}.
   
  While this is very crude, an order of magnitude estimate, the   important  point here is that the time scale $ \tau_{\rm delay}$ may dramatically vary from event to event as it strongly depends on many parameters entering the problem. In particular, it obviously depends on intensity (number of emitted particles $N$) of a previous pulse. Essentially the parameter (\ref{eq:delay1}) 
  should be interpreted as a preparation time  the   system requires  for the next eruption. 
  
  The strong variation of the $ \tau_{\rm delay}$ should be contrasted with another  parameter  which is the time duration of  a single pulse $\tau_{\rm pulse}$.
  In the AQN framework this parameter must not vary much from one event to another as it is entirely determined by internal dynamics of the AQN during the eruption    irrespective  to the intensity of the previous events, and irrespective  to the prehistory of the AQN propagation as long as condition (\ref{condition}) is met\footnote{\label{analogy2}This approximately constant parameter  $\tau_{\rm pulse}$ can be understood using the analogy with lightnings mentioned in footnote \ref{analogy1}. Indeed, the time scale between lightning flashes may dramatically vary (measured in minutes) during the same thunderstorm while the lightning strikes  themselves  are  much shorter and characterized approximately by the same duration (measured in ms).}. In different words, our proposal suggests that 
  \be
  \label{eq:tau}
 \tau_{\rm pulse}\approx \rm constant,
  \ee
  where the constant cannot be computed from first principles as it is entirely determined by the non-equilibrium dynamics of the AQN at the instant of eruption\footnote{\label{analogy3}The analogy with lightnings mentioned in footnotes \ref{analogy1} and \ref{analogy2} can be useful here: the instability in form of the runaway breakdown mechanism in lightnings which is responsible for the flashes is similar to non-equilibrium  dynamics of the AQN at the instant of eruption. Furthermore, in both cases there must be some trigger which  initiates the eruption. In case of lightning events the trigger is  thought to be related to the cosmic rays, though this  element remains a part of controversy,  see \cite{Gurevich_2001, DWYER2014147} for review. In case of the AQN eruption events such triggers could be several consequent successful   annihilation events with atmospheric material.}.  
  
  Now we are prepared  to confront the basic consequences of our  proposal (identifying  the MMEs   with the AQN annihilation events) with observations.
  
   {\it {\bf 1}. ``clustering puzzle":} 
   The multiple number of events is a very generic feature of the system as explained at the very end of subsection \ref{internal}
   as long as condition (\ref{condition}) is met. The time delays between the pulses $\tau_{\rm delay}$ dramatically fluctuate   between different events. The window for these variations is huge  according to (\ref{eq:delay1}). In fact, it could be even well outside of this window.  The time delays $\tau_{\rm delay}$ may considerably vary  even for  different events within the same cluster at the same detection point, which is consistent with observed cluster    shown on Fig. \ref{pulses}.
 Furthermore,  the AQN itself remains almost at the same location as the displacement   $ \Delta l_{\rm AQN}$  during  entire cluster of events is very tiny
   \be
  \Delta l_{\rm AQN}\sim v_{\rm AQN} \cdot  \tau_{\rm delay}\sim 20~ \rm cm,  
   \ee
   which implies that  all individual bunches  making  the cluster  are likely to  be emitted by the   AQN along the same direction, and can be recorded  and classified as MME by H10T. Each event can be viewed as an approximately  uniform front  as mere notion of the ``EAS axis" does not exist in this   framework, see also next item.  However, each individual event may appear to arrive from slightly different direction due to the inherent spread of the emitted electrons at the moment of eruption.  
   
     {\it {\bf 2}. ``particle density puzzle":} Particle density distribution $\rho(R)$  in the AQN framework is estimated by (\ref{rho}) and it shows strong fluctuation from one event to another event.  These variations are mostly related to  the intensity of the individual bursts being  expressed by the number of electrons in the bunch $N$. However, the distinct feature of the distribution in the AQN framework is that it does not depend on $R$,
     as  the notion of the ``EAS axis" does not exist in this   framework as we already mentioned. All these  generic features of the AQN framework are perfectly consistent with H10T observations as presented on Fig. \ref{density}. However, these observed features   are in  
     dramatic contradiction with   conventional EAS  prediction   shown by solid lines on Fig. \ref{density} with  different colours, depending on energy of the CR. 
          
     Furthermore, the magnitude of the density $\rho(R)$ in the AQN framework (which is mostly determined by parameter $N$)   is   much higher than  one  normally expects for CR energies $\sim (10^{17}-10^{18})$ eV. The corresponding parameter $N$ representing  the number of electrons in the bunch  was not fitted for the present studies to match the observations. Instead, it was extracted from  different experiment in dramatically different circumstances (in proposal \cite{Liang:2021rnv} to explain the ANITA anomalous events as the AQN events). 
     
       {\it {\bf 3}. ``pulse width puzzle":}  In the AQN framework the width of the pulse (\ref{eq:tau}) cannot    vary much from one event to another. It is a fundamental feature of the framework because the duration of the pulse   is entirely determined by internal dynamics of the AQN during the blast    as long as condition (\ref{condition}) is met, see footnote \ref{analogy2} with a comment.
       This feature is in perfect agreement with   observations  \cite{Beznosko:2019cI} 
  \be
   \tau_{\rm pulse}\approx (20-35) \rm ns ~~~\Leftarrow ~~~[\rm observations ]
  \ee
  for all recorded MMEs. At the same time, this feature  is in dramatic conflict with conventional picture when the duration of the pulse must depend on the distance to the EAS axis as shown by sold lines  on Fig. \ref{width}. This basic prediction of the conventional CR analysis  is due to increase of the thickness of the EAS pancake with the distance from the EAS axis.   As we already mentioned the mere notions  such as  the ``EAS axis" and the ``thickness of the EAS pancake" do not exist in the AQN   framework.
  
    {\it {\bf 4}. ``intensity puzzle":} Particle density distribution $\rho(R)$  in the AQN framework is estimated by (\ref{rho}). The corresponding event to event fluctuations  do not depend on    the distance to the EAS axis as we already mentioned. Such intensity of the events as given by  (\ref{rho}) in the AQN framework is consistent with observations shown on Fig. \ref{density}. However, the observations are  in dramatic conflict with conventional CR analysis when such huge intensity could be generated by a primary particle   with energy  well above $10^{19}$ eV with dramatically lower   event rate on the level  of once every few years. The frequency of appearance in the AQN framework is estimated in section \ref{sec:rate}, and it is consistent with observed event rate. 
    
    We conclude this section with the following comment. All our formulae  presented in this section are the order of magnitude estimations at the very best, as they include many inherent uncertainties  which are  inevitable  features of any composite  system (such as AQN) propagating in a complex environment (such as Earth atmosphere) with very large Mach number $M\equiv v_{\rm AQN}/c_s\gg 1$ where $c_s$ is speed of sound.  
    
    Nevertheless, the emergent picture  suggests  that all the {\it puzzles} formulated in Sections \ref{sec:introduction} and \ref{confronts}
    can be naturally understood  within the  AQN framework as explained above in items {\bf 1-4}. Needless to say that all phenomenological parameters used in the estimates had been fixed long ago  for dramatically different observations in different circumstances for different   environments as overviewed in Section \ref{AQN}. 
  
  \section{Conclusion and future development}\label{conclusion}
Our basic results can be summarized as follows.
We have argued that all the {\it puzzles} formulated in Sections \ref{sec:introduction} and \ref{confronts}
    can be naturally understood  within the  AQN framework as explained  in items {\bf 1-4} in previous Section, and we do not need to repeat these arguments again in this Conclusion. Instead, we   want to discuss the drastic differences between the events  induced by conventional CR showers and the AQNs. These dramatic distinct features can be tested by the  future experiments, such that our proposal  can be discriminated from any other proposals and suggestions. We list below   the following typical   features of  the AQN   events and contrast them with any other possible mechanisms which could be responsible for MMEs. 

1. The events  which are generated by the bunches of electrons as a result of eruption of the propagating  AQN in the Earth atmosphere 
suggests  an enormous number of possibilities to generate different clusters when each event within a given cluster may  have very different intensity from a previous and consequent events with very different time delays between the events. In other words,  the AQN 
proposal suggests that there should be large variety of shapes and delays between the events with very  different    patterns due to the complexity of the AQN system. It should be contrasted, for example, with hypothesis of ``delayed particles" (which was originally suggested to explain the MMEs) in which case all clusters must be the same  as they  should be determined by a specific pattern of decaying  fundamental particle  of unknown nature. 

2. A ``rule of thumb" suggests that  a typical  number of charged particles (mostly electrons and positrons) in CR air shower is $E_{\rm CR}/{\rm GeV}$, which implies that $N\approx (10^8-10^9)$ for the   energy of the primary particle  $E_{\rm CR}\approx (10^{17}-10^{18})$\,eV.   This estimate suggests that any detector which is designed to study  the EAS with energies $E_{\rm CR}\approx (10^{17}-10^{18})$\,eV are, in principle,  capable to study MMEs if the resolution of the detectors is in  $\sim 10$-ns level, similar to H10T, see also item 4 below as an alternative option to properly select and discriminate the MMEs.

3. In particular, we expect that the extension of the H10T detector would produce more multiple pulses (at each given detector) instead of simple bimodal pulses. We also expect that more detectors in the area will be recording   MMEs  because the  area covered by   each individual pulse is relatively large (few kilometres) according to (\ref{rho}), which is well above the present size of H10T instrument. 

4. A large number of charged particles $N\approx (10^8-10^9)$ in the background of the geomagnetic field ${\cal{B}}\sim 0.5$ gauss will produce the radio pulse
in both cases: the  CR-induced radio pulse \cite{Huege:2003up,HUEGE2005116} as well as AQN-induced radio pulse  \cite{Liang:2021rnv}.
However, these pulses can be easily discriminated from each other  as argued in \cite{Liang:2021rnv}. 

The main reason for the dramatic differences between these two radio pulses is that 
the AQN   event  could be viewed as an (approximately) uniform front of size $\lambda\Delta \theta\sim $km   with a constant width,   while  EAS  is characterized by central axis.  In different words, the number of particles per unit area $\rho (R)$  in the AQN case   does not depend on the distance from the central axis, in huge contrast  with conventional CR air showers when $\rho (R)$ strongly depends  the distance from the central axis.  The width of the ``pancake" in CR air shower   also  strongly depends on $R$. As a result,   the   effective number of coherent particles contributing to the radio pulse is highly sensitive to the width of the ``pancake"  when it becomes  close to the wavelength of the radio pulse.    These  distinct features lead to very different  spectral properties  of the  radio pulses  in these two cases,  which can be viewed as an independent characteristic of MMEs.  In fact, this unique feature  can be used in future  studies for purpose of discrimination and proper selection of the Multi-Modal Clustering Events.  

If future studies and tests (including the detecting   of the synchronized radio pulses with MMEs as suggested above) indeed   substantiate    our proposal it would be a strong argument supporting the AQN nature of the MMEs.

We conclude this work with the following comment. We estimated the event rates for three dramatically different puzzling CR-like events:
for mysterious TA bursts \cite{Zhitnitsky:2020shd}, for  the ANITA anomalous events   \cite{Liang:2021rnv}, and finally for the Multi-Modal Events in this work as estimated in (\ref{eq:cal N ::num}). All three puzzling phenomena are proportional  to one and the same AQN flux (\ref{Phi1}). The self-consistency between all three estimates hints on  a common nature  for these  puzzling CR-like events. 
We interpret this self-consistency in the event rates as an additional indirect argument supporting the AQN nature  for all three mysterious phenomena, while our  direct arguments  are  presented in section \ref{timing} and listed as item {\bf 1-4}.  
We finish on this optimistic note.

    \section*{Acknowledgements}
I am very thankful to Vladimir Shiltsev of FNAL who brought to my attention   the original references \cite{Beisembaev:2016cyg,2017EPJWC.14514001B,Beznosko:2019cI,2019EPJWC.20806002B,Beisembaev:2019nzd} and asked me many   questions on possible nature of MMEs, which essentially motivated and initiated this work. 
This research was supported in part by the Natural Sciences and Engineering
Research Council of Canada.

\appendix
\section{On suppression of emissivity due to the ionization}\label{sect:kappa}
The main goal of this section is to produce simple estimates of the suppression parameter $\eta (T,R)$ which enters   the cooling rate (\ref{cooling}) in case of high temperature and large ionization. It was introduced in  \cite{Ge:2020cho} as a 
suppression parameter accounting for the ionization of the AQN  after it  travelled through the Earth.
  It was also shown that $\eta (T,R)\sim 10^{-6}$ produces very reasonable results  consistent with intensity of the radiation observed by XMM-\textit{Newton} \citep{Fraser:2014wja}. 
  \exclude{
  In \cite{Ge:2020cho}, the parameter $\kappa$ was defined as:
 \begin{equation}\label{eq:nz0}
n(z)=\frac{T}{2\pi\alpha}\frac{\kappa (T)}{(z+\bar{z})^2}, ~~~~~ \kappa (T\approx 0) =1,
\end{equation}
with 
\begin{equation}\label{eq:zbar}
    \bar{z}^{-1}=\sqrt{2\pi\alpha}\cdot m_{e} \cdot \left(\frac{T}{m_{e}}\right)^{1/4}, ~~~~ n(z=0)\simeq \left(mT\right)^{3/2},
\end{equation}
where $n(z)$ is the positron number density of the electrosphere, $z=0$ corresponds to the surface of the nuggets, and $n(z=0)$ reproduces an approximate formula  for the plasma density in the Boltzmann regime at the temperature $T$ as computed in~\cite{Forbes:2008uf}. The intensity of the Bremsstrahlung radiation of the nugget is directly proportional to $n(z)$\footnote{Note that the electrosphere extends deep inside the nugget ($z<0$), but it does not contribute much to the radiation because these positrons  being far away from the Fermi surface are characterized by  very high chemical potential $\mu$.}. Since $0\le \kappa(T)\le 1$, the radiation can potentially be strongly suppressed in the limit $\kappa \rightarrow 0$. In this section we provide the physical motivation for this suppression.
}

The original computations  of the Bremsstrahlung radiation were performed in ~\cite{Forbes:2008uf} in case of low temperature and low ionization as given by (\ref{eq:P_t}). The key element of the computations was the electrosphere density  
 \begin{equation}\label{eq:nz0}
n(z)=\frac{T}{2\pi\alpha}\frac{1}{(z+\bar{z})^2},  
\end{equation}
with 
\begin{equation}\label{eq:zbar}
\nonumber
    \bar{z}^{-1}=\sqrt{2\pi\alpha}\cdot m \cdot \left(\frac{T}{m}\right)^{1/4}, ~~~~ n(z=0)\simeq \left(mT\right)^{3/2},
\end{equation}
where $n(z)$ is the positron number density of the electrosphere, $z=0$ corresponds to the surface of the nuggets, and $n(z=0)$ reproduces an approximate formula  for the plasma density in the Boltzmann regime at the temperature $T$ as computed in~\cite{Forbes:2008uf}. The 1d approximation formulated in terms of distance  $z$ from the surface   was more than sufficient sufficient to study the low temperature behaviour.  

However, after an AQN crosses  the Earth, it acquires  a very high  internal temperature of the order $T\simeq (100-200)$ keV, and approximation (\ref{eq:nz0}) is not sufficient.  
\exclude{During its journey travelling deep underground, the AQN speed $v_{\rm AQN}$ greatly exceeds the speed of sound $c_s$ by many orders of magnitude such that Mach number $M=v_{\rm AQN}/c_s\gg 1$. 
}This is because     temperature's rise will cause the electrosphere to expand well beyond the thin layer $\sim \bar{z}$ surrounding the nugget surface. Some positrons will leave the system but the majority will stay in close vicinity of the moving, negatively charged nugget core. Consequently, the nugget will acquire a negative charge of approximately $-|e|Q$ with the number of positrons $Q$ estimated as:  
\be
  \label{Q}
Q\simeq 4\pi R^2 \int^{\infty}_{0}  n(z)dz\sim \frac{4\pi R^2}{\sqrt{2\pi\alpha}}\cdot \left(m T\right)\cdot \left(\frac{T}{m}\right)^{1/4} . ~~~~~~
  \ee
  \exclude{
How many positrons precisely leave the system is not important in our argumentation and we will assume that, to first order, none of them leave the system and the negative charge of the nugget is given by Eq. \ref{Q}. In that case, when its electrosphere has expanded and the nugget core has acquired a charge Q, the nugget is said to be {\it ionized}. The ionization of the nugget refers to the charges displacement of the expanded electropshere, and not to the positrons leaving the system completely, as in the conventional atomic physics sense. 
}
The distance $\rho$ at which the positrons remain attached to the nugget is given by the capture radius $R_{\rm cap}(T)$, determined by the Coulomb attraction:
\be
\label{capture}
\frac{\alpha Q(\rho)}{\rho} > \frac{m v^2}{2}\approx T ~~~{\rm for}~~~ \rho\lesssim  R_{\rm cap}(T).
\ee
\exclude{
When the AQNs exit the Earth's atmosphere their temperature will slowly decreases with the rate computed in paper-I. The positrons will then return to their original positions as the AQNs are moving away in empty space and eventually assume the temperature $T\approx 0$ far away from Earth. 

From these arguments it is expected that the positron number density at the moment of exit of the Earth surface will be much smaller than at the later times, i.e. $n(r\simeq R_{\oplus}) \ll n(r\gg R_{\oplus})$, because the expansion of the electrosphere can be quite significant, even if only a small number of positron effectively leave the system.
Therefore, the Bremsstrahlung emission of the AQN (proportional to the number density square), immediately at the Earth exit point is strongly suppressed in comparison to the late times, when the nugget has reached a distance $r\sim 8R_{\oplus}$, where XMM-\textit{Newton} is operational which corresponds to the computations done in paper-I.
}

In order to estimate $\eta$ entering  (\ref{cooling}) we start our analysis with an estimation of the positron density $n(\rho, T)$ when the AQNs enter the   Earth atmosphere moving in upward direction. Formula (\ref{capture}) shows that the capture radius $R_{\rm cap}(T)$ could be  much  larger than $R$, i.e. we have $ R\lesssim \rho\lesssim R_{\rm cap}$. The expression (\ref{eq:nz0}), which is valid when $z=|\rho-R|\ll R$, in close vicinity to the core, breaks down for $\rho\gtrsim  R$. In that case the curvature of the nugget surface cannot be neglected and one should use a truly 3-dimensional formula for $n(\rho, T)$ instead of the 1-dimensional approximation (\ref{eq:nz0}). 
We will assume that the density $ n(\rho, T)$ behaves as a power law for $\rho\gtrsim  R$ with exponent $p$:  
  \be
\label{eq:density}
n(\rho, T)\simeq n_0(T) \left(\frac{R}{\rho}\right)^p, ~~~ \rho\gtrsim  R,
\ee
where $n_0 (T)\equiv n(\rho=R, T)$ is a normalization factor and $p$ a free parameter. 
Eq. (\ref{eq:density})  is consistent with our previous numerical studies \cite{Forbes:2009wg} of the electrosphere with $p\simeq 6$. It is also consistent with the Big Bang Nucleosyntheses (BBN) study discussed in \cite{Flambaum:2018ohm} and with the conventional Thomas-Fermi model\footnote{In \cite{Landau}, the dimensionless function $\chi (x) $ behaves as $\chi\sim x^{-3}$ at large $x$. The  potential $\phi=\chi (x)/x$ behaves as 
  $\phi\sim x^{-4}$. The density of electrons in Thomas-Fermi model scales as $n\sim \phi^{3/2}\sim x^{-6}$ at large $x$.}  at $T=0$ \citep{Landau}. 
While keeping $p$ as a free parameter, we will show below that our main claim is not very sensitive to the precise value of $p$. The normalization $n_0 (T)$ can be estimated from the condition that the finite portion of the positrons satisfying Eq. (\ref{capture}) is such that the total number of positrons surrounding the nugget is approximately equal to the ionization charge $Q$ determined by (\ref{Q}). Therefore, we arrive to the following expression for the  normalization factor 
  $n_0 (T)$:
  \be
  \label{normalization}
  4\pi \int_R^{\infty} \rho^2 d\rho~ n(\rho, T) \simeq Q (T), 
    \ee
Note that the integrant $\rho^2 n(\rho, T)$ is mostly dominated by the inner shells $\rho\sim R$ such that the intgration can be extended to infinity with very high accuracy, instead of cutting off at $R_{\rm cap}$.
  The resulting estimate for $n_0(T)$  assumes the form:
 \be
 \label{n_0}
  n_0(T)\simeq \frac{ \left(m T\right)}{\sqrt{2\pi\alpha}}\cdot \frac{(p-3)}{R}\cdot \left(\frac{T}{m}\right)^{1/4},
   \ee 
  which, as expected, is smaller than $n(z=0)\simeq \left(mT\right)^{3/2}$ because the positrons now are distributed over a distance of order $R$ from the nugget core surface, rather than over a distance of order $\bar{z}$. This suppression is convenient to represent in terms of the dimensionless ratio:
  \be
  \label{eq:suppression}
  \frac{n_0(T)}{\left(m T\right)^{3/2}}\simeq \frac{ (p-3)}{\sqrt{2\pi\alpha}}\cdot \frac{1}{(mR)}\cdot \left(\frac{m}{T}\right)^{1/4} 
  \ll 1.
  \ee
 
The next step is the calculation of the Bremsstrahlung emissivity by the nuggets due to the  strong suppression of the plasma density (\ref{eq:suppression}) as a result of  the electrosphere's expansion.
The spectral surface emissivity is denoted as $dF/d\omega=dE/dtdAd\omega$, representing the energy emitted by a single nugget per unit time, per unit area of nugget surface and per unit frequency. For low temperature, when approximation (\ref{eq:nz0}) for the flat geometry is justified, the corresponding expression assumes the form~\citep{Forbes:2008uf}:
\begin{equation}\label{eq:dFdomega0}
    \frac{dF}{d\omega}(\omega, T)=\frac{1}{2}\int_{0}^{\infty}dz ~n^2 (z)~{\cal{K}}(\omega, T)
\end{equation}
where $n(z)$ is the local density of positrons at distance $z$ from quark nugget surface and 
  function ${\cal{K}}(\omega)$ does not depend on $z$ and describes the spectral dependence of the system:
\begin{equation}
\label{eq:dQ0}
    {\cal{K}}(\omega, T)= \frac{4\alpha}{15}\left(\frac{\alpha}{m}\right)^2 2\sqrt{\frac{2T}{m\pi}} \left(1+\frac{\omega}{T}\right){\rm e}^{-\omega/T} h\left(\frac{\omega}{T}\right). 
\end{equation}
The dimensionless function $h\left(\frac{\omega}{T}\right)$ in (\ref{eq:dQ0}) is a slowly varying logarithmic function of ${\omega}/{T}$ which  was  explicitly computed in ~\cite{Forbes:2008uf}. In order to calculate the emissivity with the positron density (\ref{eq:density}), when the nugget core is {\it ionized}, we have to replace the integral $\int dz$ in (\ref{eq:dFdomega0}) by the spherical integration over $\rho$:
\be
\label{replacement}
4\pi R^2 \int_{0}^{\infty}dz ~n^2 (z)~~ \Rightarrow ~ 4\pi n^2_0(T)\int_R^{\infty}\rho^2 d\rho \left(\frac{R}{\rho}\right)^{2p}~~~
\ee
where the normalization $n_0(T)$ is determined by (\ref{n_0}). Using the positrons distributed according to Eq. (\ref{eq:density}) the nugget surface emissivity can be calculated as follows:
\be
\label{emissivity1}
 \frac{dF}{d\omega}(\omega, T)= \frac{{\cal{K}}(\omega, T)}{2R}  \frac{(mT)^2}{2\pi\alpha}  \left[\frac{(p-3)^2}{2p-3}\right] \left(\frac{T}{m}\right)^{\frac{1}{2}}.~~~~~~~
\ee
The key result here is the strong suppression factor $R^{-1}$ which was not present in the no-ionization case (\ref{eq:dFdomega0}) when the positrons are localized in a thin layer. It is instructive to compare the emissivity given by eq. (\ref{emissivity1}) accounting for ionization of the nugget core with the original formula (\ref{eq:dFdomega0}). The dimensionless  ratio is given by
\be
\label{ratio1}
\eta\equiv\frac{\frac{dF}{d\omega}^{(\rm ion)}}{{ \frac{dF}{d\omega}^{(\rm no~ion)}}}=\frac{\left[\frac{(p-3)^2}{2p-3}\right]}{3(mR)}\frac{1}{\sqrt{2\pi\alpha}} \left(\frac{m}{T}\right)^{\frac{1}{4}}\sim 10^{-6}. ~~~~~~~~
\ee
\exclude{
This estimate assumes that all weakly coupled positrons contribute to the charge Q given by eq. (\ref{Q}). Therefore, Eq.  (\ref{ratio1}) is the lower limit (strongest possible) of suppression. In practice, the suppression could be less dramatic because a small portion of the positrons could remain in thin layer surrounding the nugget's surface and contribute to the emission with no suppression factor. 
}The total emissivity integrated over all frequencies in case of low temperature without ionization has been computed in  ~\citep{Forbes:2008uf} and it is given by (\ref{eq:P_t}).
\exclude{
\be
\label{total-Forbes}
  F^{(\rm no~ion)}_{\rm tot} (T)= \int d \omega  \frac{dF}{d\omega}(\omega, T)\simeq \frac{16}{3} \cdot\frac{T^4 \alpha^{\frac{5}{2}}}{\pi} \cdot \left(\frac{T}{m}\right)^{\frac{1}{4}}.
\ee
}
In case of strong ionization, Eq. (\ref{ratio1}) implies:
 \be
\label{total}
  F^{(\rm ion)}_{\rm tot} (T)=  \eta (T, R) \cdot  F^{(\rm no~ion)}_{\rm tot} (T).
\ee
The critical element leading to the suppression factor (\ref{ratio1}) is the very small quantity $(mR)^{-1}\sim 10^{-6}$. It is a direct result of the expanding electrosphere and high level of ionization at very high  temperature $T$. 

The expression (\ref{total}) with numerical suppression   (\ref{ratio1}) is precisely the formula (\ref{cooling}) which we used in main body of  the text to proceed with our key  arguments. 

\exclude{
Since the emissivity is proportional to $n^2(z)$ according to  (\ref{eq:dFdomega0}), it implies that $\eta\sim \kappa^2$. Consequently, the empirical suppression factor $\kappa$ entering (\ref{eq:nz0}) is of order $\kappa\sim 10^{-3}$ being consistent  with our phenomenological analysis presented in paper-I. 
It is important to emphasize that the ratio  $\eta$ is not very sensitive to our assumption on value of the exponent $p$ within a reasonable window. In particular, for $p\simeq 6$ the expression in the brackets   equals to unity. It is also  important to emphasize that  $\eta (T, R) $  does not depend on the frequency $\omega$
such that the computation of the spectrum performed in paper-I is perfectly justified\footnote{this claim does include the region of very small frequencies $\omega \ll T$ which was cutoff in paper-I.  Small frequency region does not play any  role in the  present work.}. 
}


\section{On $e^+e^-$ emission at high temperature   in high density QCD phases.}\label{pair-production}
The main goal of this Appendix is to overview the photon emission and the $e^+e^-$  production at  high temperature $T\gtrsim 10^2$ keV from high density QCD phases. The corresponding studies \cite{Usov:1997ff,Usov:2001sw, PhysRevD.70.023004,Harko_2005,Caron:2009zi,Zakharov:2010ch,Zakharov:2010yz}
 have been carried out in the past  in context of the quark stars. In context of the present work all the key ingredients relevant for $e^+e^-$  production are also present in the system. Indeed,  the AQN is characterized by very high temperature $T\gtrsim 10^2$ keV, the quark core is assumed to be in CS dense phase, and a strong internal electric field is also present in the system. However, we cannot literally use the results from  the previous studies obtained in context of the quark stars because the size of the AQN is much smaller than the  relevant mean free paths  for  all elementary processes  as discussed in details in   \cite{Liang:2021rnv}.
 As a result the thermal equilibration cannot be achieved  in the AQN system, and entire physics is determined by non-equilibrium dynamics in high temperature regime.  It should be contrasted with large size  quark stars where the thermal equilibrium is maintained. 

Nevertheless, it is very instructive to review the relevant results from the previous  studies  \cite{Usov:1997ff,Usov:2001sw, PhysRevD.70.023004,Harko_2005,Caron:2009zi,Zakharov:2010ch,Zakharov:2010yz} on quark stars due to the following reasons. First, it explicitly shows
the role of the main ingredients of the system, such as temperature and the large electric field. Secondly, it demonstrates   the  
 complexity of the problem when even a much simpler case of the bare quark star remains to be  a matter of debates. 

The idea on possibility of the $e^+e^-$ emission at high temperature from quark stars was originally suggested in \cite{Usov:1997ff,Usov:2001sw}.
The temperature range considered in \cite{Usov:1997ff,Usov:2001sw} includes the typical temperatures $T\gtrsim 10^2$ keV which is expected to occur in our case when the AQN exits the Earth's surface as mentioned in Section \ref{internal}. In refs. \cite{PhysRevD.70.023004,Harko_2005} the authors argue that 
bremsstrahlung radiation from the electrosphere could be much more important than $e^+e^-$ emission. It has been also argued that a number of effects such as 
the boundary effects, inhomogeneity of the electric field, and the Landau-Pomeranchuk-Migdal (LPM) suppression may dramatically modify the emission rate. 
In \cite{Caron:2009zi} it has been argued that the Pauli blocking 
will strongly suppress the  bremsstrahlung emission. Finally, in refs. \cite{Zakharov:2010ch,Zakharov:2010yz} it has been argued that the so-called mean field bremsstrahlung could be the dominant mechanism. 

It is not the goal of the present work to critically analyze all these suggested mechanisms of the emission.
Rather, our goal is to demonstrate that even a relatively simple system of the bare quark star remains to be  a matter of debates. Our case of the emission from AQN at high temperature is even more complicated as it is determined by non-equilibrium dynamics.  

In the present work  we do not even attempt to solve this very ambitious problem of estimating the absolute intensity rate of the $e^+e^-$ and $\gamma$ emissions.
Rather, the absolute number $N\simeq (10^8-10^9)$     of the $e^+e^-$ pairs  in a single pulse was extracted from ANITA anomalous events as explained in Section \ref{timing}.
In particular, this number $N$ enters the estimate (\ref{rho}) for the density of the observed particles by H10T experiment \cite{Beisembaev:2016cyg,2017EPJWC.14514001B,Beznosko:2019cI,2019EPJWC.20806002B}.  Our goal here is much less ambitious as we try to estimate  the relative ratio $r$ between the  $e^+e^-$ pair production and the $\gamma$ radiation  assuming that the emission  of  both components is  mostly originated from the region in electrosphere
where the Boltzmann regime is justified, and  the Pauli blocking suppression is not very  dramatic and can be ignored. 

   In this case, the relative ratio $r$ between these two components contributing to the emission can be estimated as  follows. It is assumed  that the dominant contribution to the $\gamma$ emission comes from  
  the Bremsstrahlung radiation ($e^+e^+\rightarrow e^+e^+ +\gamma$) resembling   the low temperature case  considered in ~\cite{Forbes:2008uf}. 
  The $e^+e^-$  pair  is produced  by a similar mechanism 
 through the virtual photon ($e^+e^+\rightarrow e^+e^+ +\gamma^*$) which consequently decays to the pair: $\gamma^*\rightarrow e^+e^-$.  If this is the dominant mechanism of radiation the pair production  is expected to be suppressed by  a  factor ${\alpha}/{2\pi}$   as a result of  conversion of the virtual photon $\gamma^*\rightarrow e^+e^-$ to pair\footnote{The cross section for pair production as a result of collisions of two particles  is known exactly\cite{landau-QED}. It contains, of course factor ${\alpha}/{2\pi}$  entering (\ref{ratio-Appendix})  along with many other numerical factors reflecting a complex  kinematics of the process.}. It must be also suppressed by a factor $  \exp(-{2m}/{T} )$ describing  the suppression in the density distribution because  the virtual  photons must have sufficient energy $\gtrsim 2m$  to  produce $e^+e^-$ pair. This oversimplified estimate leads to the following expression
  \be
  \label{ratio-Appendix}
 r\equiv \frac{\la E_{e^+e^-}\ra}{\la E_{\gamma}\ra}\sim \frac{\alpha}{2\pi}\exp \left(-\frac{2m}{T}\right)\sim 0.7\cdot 10^{-5},  
   \ee
     where for numerical estimates we use $T\simeq 200$ keV, and ignored a possible complicated function  which could  depend on  parameter $T/m\sim 0.4$, which is order of unity for our case. 
  We must emphasize that  (\ref{ratio-Appendix})  is really an order of magnitude estimate as mechanism of emission is not a thermodynamically equilibrium process as we  already previously mentioned. Formula   (\ref{ratio-Appendix})  is used in the main body of the text in (\ref{ratio}) for our estimate
     (\ref{eq:delay1}) for $ \tau_{\rm delay}$ which is measured by H10T experiment \cite{Beisembaev:2016cyg,2017EPJWC.14514001B,Beznosko:2019cI,2019EPJWC.20806002B} as it describes a typical time delay between subsequent  pulses representing the Multi-Modal Clustering  Events.

\bibliography{MM}

\end{document}